\documentclass[twocolumn,tighten,twocolappendix]{aastex63}
\pdfoutput=1 
\usepackage{amsmath,amstext}
\usepackage[T1]{fontenc}
\usepackage{ae,aecompl}
\usepackage[utf8]{inputenc}
\usepackage{apjfonts} 
\usepackage[figure,figure*]{hypcap}
\usepackage[ruled,vlined]{algorithm2e}
\usepackage{enumitem}
\usepackage{bm}
\usepackage{pifont}
\input{hyperlink-year-only-natbib-patch}

\newcommand*{\http}[1]{\href{http://#1}{#1}}
\newcommand*{\https}[1]{\href{https://#1}{#1}}

\graphicspath{{./}{figures/}}

\shorttitle{Gaia Wavelet Dwarf Galaxy Search}
\shortauthors{Darragh-Ford et al.}

\begin{document}

\title{Searching for Dwarf Galaxies in \emph{Gaia} DR2 Phase-Space Data Using Wavelet Transforms}

\correspondingauthor{}
\email{edarragh@stanford.edu}
\author[0000-0002-8800-5652]{Elise Darragh-Ford}
\author[0000-0002-1182-3825]{Ethan~O.~Nadler}
\author[0000-0002-7042-6424]{Sean McLaughlin}
\author[0000-0003-2229-011X]{Risa~H.~Wechsler}
\affiliation{Kavli Institute for Particle Astrophysics and Cosmology and Department of Physics, Stanford University, Stanford, CA 94305, USA}
\affiliation{SLAC National Accelerator Laboratory, Menlo Park, CA 94025, USA}

\begin{abstract}
We present a wavelet-based algorithm to identify dwarf galaxies in the Milky Way in \emph{Gaia} DR2 data. Our algorithm detects overdensities in 4D position--proper motion space, making it the first search to explicitly use velocity information to search for dwarf galaxy candidates. We optimize our algorithm and quantify its performance by searching for mock dwarfs injected into \emph{Gaia} DR2 data and for known Milky Way satellite galaxies. Comparing our results with previous photometric searches, we find that our search is sensitive to undiscovered systems at Galactic latitudes~$\lvert b\rvert>20^{\circ}$ and with half-light radii larger than the 50\% detection efficiency threshold for Pan-STARRS1 (PS1) at (\emph{i}) absolute magnitudes of =$-7<M_V<-3$ and distances of $32\ \mathrm{kpc} < D < 64\ \mathrm{kpc}$, and (\emph{ii}) $M_V< -4$ and $64\ \mathrm{kpc} < D < 128\ \mathrm{kpc}$. Based on these results, we predict that our search is expected to discover $5 \pm 2$ new satellite galaxies: four in the PS1 footprint and one outside the Dark Energy Survey and PS1 footprints. We apply our algorithm to the \emph{Gaia} DR2 dataset and recover $\sim 830$ high-significance candidates, out of which we identify a ``gold standard'' list of $\sim 200$ candidates based on cross-matching with potential candidates identified in a preliminary search using \emph{Gaia} EDR3 data. All of our candidate lists are publicly distributed for future follow-up studies. We show that improvements in astrometric measurements provided by \emph{Gaia} EDR3 increase the sensitivity of this technique; we plan to continue to refine our candidate list using future data releases.

\end{abstract}

\keywords{\href{http://astrothesaurus.org/uat/80}{Astrometry (80)}; \href{http://astrothesaurus.org/uat/1295}{Proper motions (1295)}; \href{http://astrothesaurus.org/uat/416}{Dwarf galaxies (416)}; \href{http://astrothesaurus.org/uat/1965}{Computational methods (1965)}}

\section{Introduction} \label{sec:intro}
Our understanding of the structures that make up the Milky Way (MW) and its local environment is a cornerstone of modern astrophysics and near-field cosmology. Mapping out these structures has provided insights into the assembly history and growth of the Galactic halo (e.g.\ \citealt{Bullock_2001}), the physics of galaxy formation (e.g.\ \citealt{Mashchenko_2008,Helmi_2010,Graus_2019}), and even into fundamental outstanding questions in cosmology including the nature of dark matter \citep{Bullock2017}. This is because observations of the structures composing and surrounding the MW allow us to observe the luminous tracers of extremely low-mass dark matter halos, including the stellar remnants of disrupted systems in the Galactic halo. 
The abundance, structure, and stellar components of these low-mass halos are in turn deeply related to the microphysics of a given dark matter model. Thus, ultra-faint dwarf galaxies (UFDGs), defined as objects with luminosities $L < 10^5\  \mathrm{L}_\odot$ \citep{Simon_2019}, are an exciting tool for understanding the properties of dark matter. Analyses using the currently known population of UFDGs have already placed novel constraints on a range of popular dark matter scenarios (e.g.\ \citealt{Macci__2010,Vogelsberger_2012,Nadler_2019,nadler2020milky}). However, this type of analysis is still limited by the relatively small number ($\sim 60$) of Milky Way satellites that have been detected to date, an even smaller number of which have been spectroscopically verified as dark matter-dominated systems \citep{Geha2009,Willman_2012, Kirby_2013,Drlica-Wagner2020}.  While cold dark matter theory predicts a large abundance of UFDGs down to the star-formation limit for dark matter halos, detecting the very faint galaxies that reside in these low-mass objects is extremely challenging.

So far, wide-field photometric surveys like the Sloan Digital Sky Survey (SDSS), the Dark Energy Survey (DES) and Pan-STARRS1 (PS1) have been most successful in increasing the number of known objects in this regime (e.g.\ \citealt{Willman2005,Belokurov2006,Zucker2006, Laevens2015,Bechtol150302584,Drlica-Wagner150803622, Drlica-Wagner2020}). These searches tend to rely on filtering stars by their color and brightness and then smoothing stellar fields spatially to detect statistical overdensities of individually resolved stars. However, as these dwarf searches push to greater distances and fainter surface brightnesses, this traditional approach is hampered by issues like robust star--galaxy separation. Furthermore, this approach does not make use of the fact that stars within dwarf galaxies are expected to have both coherent positions and velocities. As we will demonstrate in this work, astrometric measurements that access this velocity information can be used to develop a unique complementary approach to detecting UFDGs.

\emph{Gaia} \citep{Gaia_mission,Gaia180409365} is a space-based mission by the European Space Agency that has delivered unparalleled astrometric measurements for over a billion stars across the entire sky. This data has already provided invaluable insights into our understanding of the interrelated properties of the Milky Way stellar halo, density profile, and accretion history (e.g.\ \citealt{Belokurov180203414,Deason180510288,Helmi180606038,Cautun191104557}; and see the review in \citealt{Helmi200204340}). Additionally, kinematic data from \emph{Gaia} has been used to broaden our understanding of the orbital and internal characteristics of known dwarf galaxies \citep[e.g.][]{Simon180410230,Fritz180500908} and \emph{Gaia} proper motion measurements have also helped refine recent photometric searches for UFDGs. For instance, \cite{Mau191203301} used kinematic information to help distinguish between candidate UFDGs and chance associations of stars.

Perhaps even more excitingly, \emph{Gaia} data has already been used to detect previously unknown dwarf galaxies. In \cite{Torrealba181104082}, \emph{Gaia} astrometric and photometric data combined with identified overdensities of RR Lyrae stars were used to discover Antlia 2, an extremely diffuse dwarf satellite galaxy near the Milky Way's disk. While deeper Dark Energy Camera imaging data was required for confirmation, this discovery highlights the power of the 
\emph{Gaia} dataset, which makes up for limited depth with excellent star--galaxy separation and very accurate and precise astrometric information. \emph{Gaia} measurements will only continue to improve with the upcoming DR3 data release, which is expected to significantly increase the precision of proper motion and parallax measurements. This makes comprehensive searches for undetected substructure that incorporate the unique strengths of the \emph{Gaia} dataset a very promising and thus far under-explored avenue of research.

From the analysis side, machine-learning based approaches have recently been applied to \emph{Gaia} data to disentangle the various structures that make up the Milky Way halo, with promising results. For example, deep neural networks have been used to separate accreted and in-situ stars in the \emph{Gaia} DR2 dataset \citep{Ostdiek190706652}, and clustering algorithms like DBSCAN \citep{Koppelman_2019}, $k$-means \citep{Mackereth2018}, and Gaussian mixture models \citep{Necib190707681} have been used to identify dynamically distinct structures in the accreted halo. 

In contrast, wavelet analysis as a tool for image processing has been around for decades \citep{xiong1997deblocking} and has been used to understand Milky Way substructure for almost as long \citep{Chereul_1998}. However, the technique has recently seen a renewed interest in the astronomy community. The wavelet transform can be thought of as functioning similarly to a localized Fourier transform, and thus can be tuned to highlight structure at specific scales in an image. \cite{Antoja150704353} used this property of the transform to identify overdensities corresponding to mock UFDGs in synthetic \emph{Gaia} DR2 data by performing independent transforms on the position and proper motion planes. 

In this work, we expand upon this idea by performing a wavelet transform on the full 4D distribution of stars in position--proper motion space to amplify coherent phase-space overdensities, rather than trying to cross-match peaks detected in 2D position and proper motion space separately. We show that this method is able to recover mock dwarf galaxies injected into the \emph{Gaia} DR2 data over a large fraction of the sky with better sensitivity than previous dwarf galaxy searches have achieved. Specifically, we predict that our search is sensitive to undiscovered systems at Galactic latitudes~$\lvert b \rvert>20^{\circ}$ over the $\sim 3/4$ of the sky covered by PS1 with 
(\emph{i}) absolute magnitudes of $-7<M_V<-3$ and distances of $32\ \mathrm{kpc} < D < 64\ \mathrm{kpc}$, and (\emph{ii}) absolute magnitudes of $M_V< -4$ and distances of $64\ \mathrm{kpc} < D < 128\ \mathrm{kpc}$.
We use these sensitivity predictions combined with the galaxy--halo connection model from \cite{Nadler191203303} to estimate the number of undiscovered dwarfs expected to be recovered by this analysis. We estimate that $5 \pm 2$ undiscovered systems should be detected, four in the PS1 footprint and one in the region of $\lvert b\rvert>20^{\circ}$ sky that is outside both DES and PS1.  
By applying our search algorithm to the full \emph{Gaia} dataset, we have compiled a list of potential dwarf candidates, which is made publicly available online for follow-up studies.\footnote{\href{https://dwarfswaves.github.io}{https://dwarfswaves.github.io}}

This paper is structured as follows. In Section \ref{sec:data}, we describe the \emph{Gaia} data used in this analysis. In Section \ref{sec:algorithm}, we introduce the discrete wavelet transform, the details of our dwarf search algorithm, and our method for measuring candidate significance. In Section \ref{sec:analysis}, we calibrate the significance of candidates returned by the algorithm by searching for mock dwarf galaxies injected into \emph{Gaia} data (Section \ref{sec:mock_analysis}), estimating the sensitivity of our search for these mock dwarfs (Section \ref{sec:sensitivity}), and comparing to search results using the sample of known Milky Way satellite galaxies (Section \ref{sec:dwarf_analysis}). We present the results of our full-sky search in Section \ref{sec:full_sky_analysis}, which we publicly distribute as a list of high-significance candidates for future follow-up studies. We conclude and highlight avenues for future work in Section \ref{sec:conclusion}.




\section{Data}
\label{sec:data}

We make use of the full \emph{Gaia} DR2 dataset for sources with measured position, proper motion, and parallax. This consists of 1332 million individual sources \citep{Lindegren_2018}. In order to remove foreground stars, we impose the following cuts: 
\begin{align}
    &-5\ \mathrm{mas\ yr}^{-1} < \mu_{\alpha} < 5\ \mathrm{mas\ yr}^{-1},&\label{eq:mua_cut} \\ 
    &-5\ \mathrm{mas\ yr}^{-1} < \mu_{\delta} < 5\ \mathrm{mas\ yr}^{-1},&\label{eq:mudelta_cut} \\ 
    &\varpi - 3\sigma_\varpi < 0.0\ \mathrm{mas},&\label{eq:parallax_cut}
\end{align}
where $\mu_{\alpha}$ ($\mu_{\delta}$) is the $\alpha_{2000}$ ($\delta_{2000}$) proper motion component. The cut in parallax ($\varpi$) is included to filter out foreground stars by removing objects with robustly measured parallaxes. The proper motion cuts are imposed to ensure that we are sensitive to the range of proper motion values that have been measured for known UFDGs (e.g.\ \citealt{Gaia_kinematics,Simon180410230,Fritz180500908}) and to ensure that the dwarfs have roughly similar sizes in position and proper motion space when the position and proper motion fields are converted to a 4D histogram as described in Section \ref{sec:algorithm}. We can estimate the approximate size of a UFDG in position and proper motion space following \cite{Antoja150704353}, 
\begin{align}
  & \Delta\theta \sim 0.0573~\frac{r_{1/2}}{D},\label{eq:deltatheta} & \\ 
  & \Delta \mu \sim 0.211~\frac{\sigma_v}{D},\label{eq:deltamu}&
\end{align}
where $\Delta\theta$ is the angular extent of the UFDG (in degrees) and $\Delta\mu$ is its extent in proper motion space (in $\mathrm{mas\ yr}^{-1}$), $r_{1/2}$ is its projected, azimuthally averaged half-light radius (in $\mathrm{pc}$), $D$ is its heliocentric distance (in $\mathrm{kpc}$), and $\sigma_v$ is its velocity dispersion (in $\mathrm{km\ s}^{-1}$).

For the dwarf galaxy Tucana II ($\sigma_v = 8.6^{+4.4}_{-2.7}\ \mathrm{km\ s}^{-1}$, $r_{1/2} = 165\ \mathrm{pc}$, $D = 58\ \mathrm{kpc}$; \citealt{Walker_2016}), which is representative of the undiscovered systems our search is sensitive to, Equations \ref{eq:deltatheta}--\ref{eq:deltamu} yield $\Delta\theta = 0.16^\circ$ and $\Delta \mu = 0.03 \  \mathrm{mas\ yr}^{-1}$. For a field of approximately $2^\circ$ across in position and $10\ \mathrm{mas\ yr}^{-1}$ in proper motion binned into a $96\times96\times96\times96$ square histogram, we expect Tucana II to span approximately seven pixels in position space and a pixel or two in proper motion space. These roughly equal sizes simplify our analysis because they allow us to apply the same wavelet transforms to both the position and proper motion dimensions. 

Fields are generated using an optimal packing of points on a sphere. A total of 33,002 points are chosen corresponding to a minimum distance between points, of approximately 1 degree. For each of those points a field is then generated using a gnomonic projection of the sphere centered at that point. The field is generated using a fixed gnomonic width that corresponds to $\pm 1\ \mathrm{deg}$ in $\delta_{2000}$. Thus, there is approximately $1\ \mathrm{deg}^2$ overlap between adjacent fields ensuring that objects on the edges of fields are not missed due to edge effects in the wavelet analysis. Gnomonic projections are computed via
\begin{align}
    & x = \frac{\cos{\phi}\sin{(\lambda-\lambda_0})}{\cos{c}}\label{eq:gnomx} & \\ 
    & y = \frac{\cos{\phi_1}\sin{\phi} - \sin{\phi_1}\cos{\phi}\cos{(\lambda-\lambda_0})}{\cos{c}}, \label{eq:gnomy}& 
\end{align}
where
\begin{equation}
   \cos{c} = \sin{\phi_1}\sin{\phi} + \cos{\phi_1}\cos{\phi}\cos{\lambda-\lambda_0} 
\end{equation}
and $\lambda_0$ and $\phi_1$ are the coordinate centers of the projection in radians.

 The first slice in $\alpha_{2000} (\delta_{2000})$ is recentered by $\alpha_{{2000}_j} = 0 + \frac{w_\alpha}{2} (\delta_{{2000}_j} = -90 + \frac{w_\delta}{2})$ and the final slice in $\alpha_{2000} (\delta_{2000})$ is recentered by $\alpha_{{2000}_j} = 360 - \frac{w_\alpha}{2} (\delta_{{2000}_j} = 90 - \frac{w_\delta}{2})$; for all the other fields, the centers remain the same. \emph{Gaia} stars are then selected that correspond to each of these fields.

\begin{figure*}[htb!]
\centering
\includegraphics[width=\textwidth]{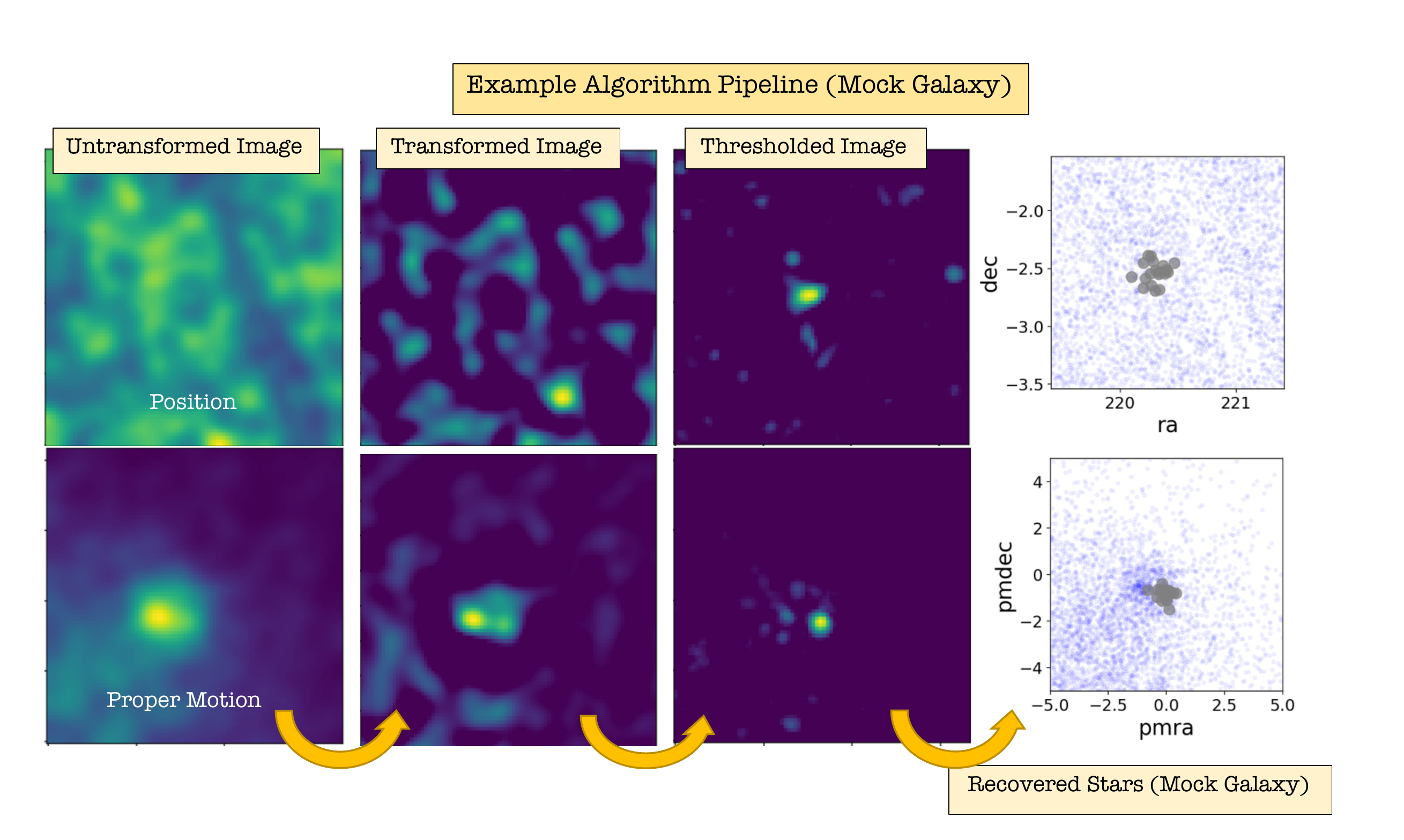}
\caption{Example of our wavelet search algorithm for a mock dwarf galaxy injected into \emph{Gaia} DR2 data. \emph{First column:} The histogram of \emph{Gaia} stars passing our proper motion and parallax cuts (Equations \ref{eq:mua_cut}--\ref{eq:parallax_cut}) before applying the wavelet transform, smoothed using a Gaussian filter and projected into position space (\emph{top}) and proper motion space (\emph{bottom}). \emph{Second column:} The transformed histogram of \emph{Gaia} stars after normalizing the \emph{Gaia} field to a corresponding random field with the same Poisson rate according to Equation \ref{eq:sigmaw}. \emph{Third column:} The transformed histogram after thresholding on normalized overdensities with Poisson significances greater than $\sim 5\sigma$. The left three histograms have all been normalized to take values in $[0,1]$. \emph{Fourth column:} The position and proper motion of the stars returned by our algorithm. The grey points show the stars returned by our algorithm for this field, with completeness and purity values indicated in Figure \ref{fig:example_field}. The corrected wavelet significance of this overdensity is $\sigma_w$ = 11.0 The dwarf in this image has $M_V = -4.3$ and $r_{1/2} = 145\ \mathrm{pc}$, and is at a distance of $D = 57\ \mathrm{kpc}$. This mock galaxy was chosen to approximately match the properties of Tucana II; we illustrate our algorithm's performance on Tucana II in Figure \ref{fig:analysis_pipeline_tuc}.}
\label{fig:analysis_pipeline}
\end{figure*}

\begin{figure*}[htb!]
\centering
\includegraphics[width=0.75\textwidth]{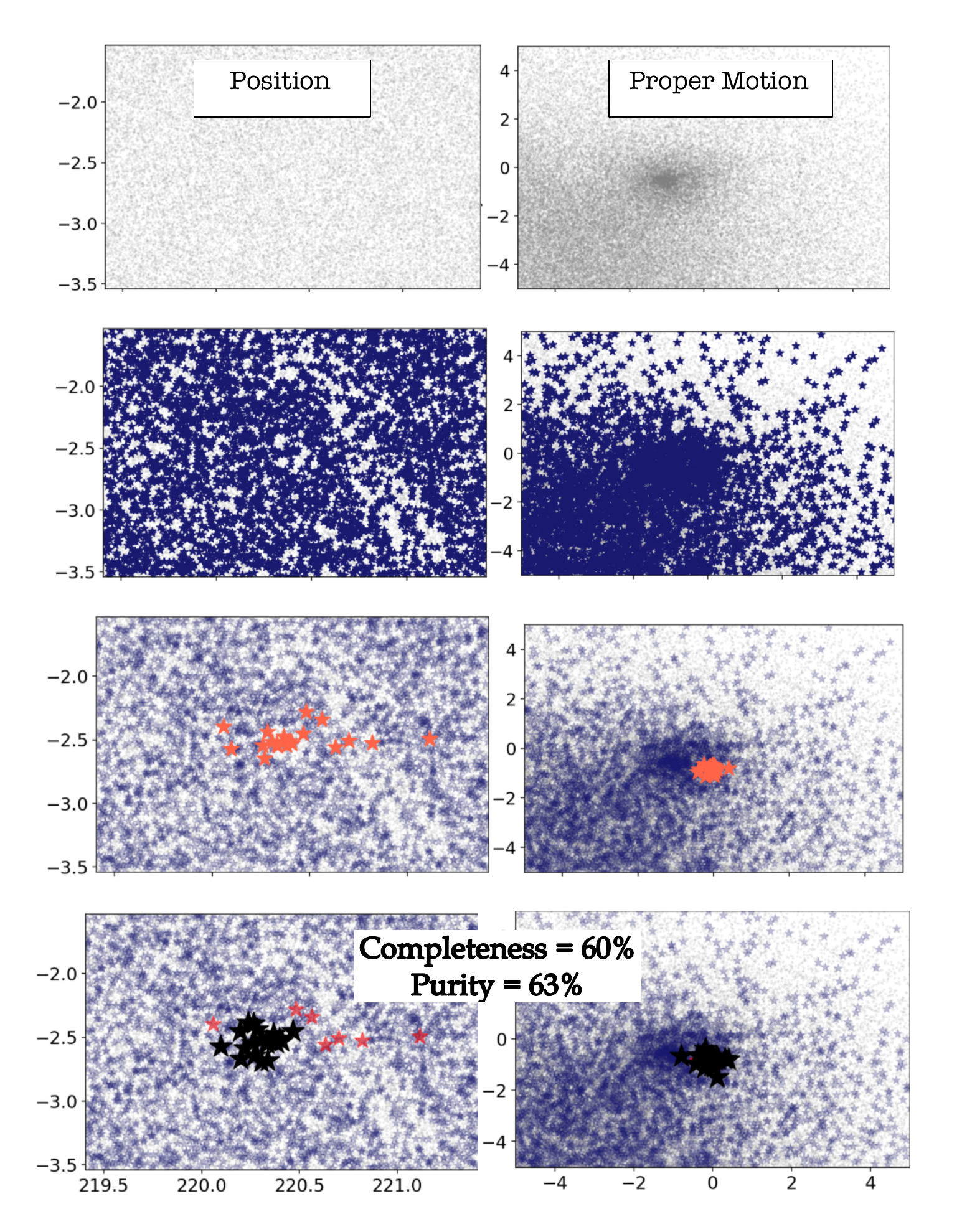}

\caption{Example position (\emph{left}) and proper motion (\emph{right}) fields for the same mock galaxy shown in Figure \ref{fig:analysis_pipeline}, inserted into \emph{Gaia} DR2 data. \emph{Row 1}: All \emph{Gaia} stars in the selected field with no parallax cut. \emph{Row 2}: Stars shown in blue pass our parallax cut of $\varpi - 3\sigma_\varpi < 0.0\ \mathrm{mas}$. \emph{Row 3}: Stars shown in red belong to the injected mock dwarf galaxy. \emph{Row 4}: Stars shown in black are those ultimately recovered by our wavelet search algorithm. The mock dwarf in this image has $M_V = -4.3$ and $r_{1/2} = 145\ \mathrm{pc}$, and is at a distance of $D = 57\ \mathrm{kpc}$. $60\%$ of the stars in the dwarf are recovered by our search algorithm, and $63\%$ of the stars recovered by the algorithm are actually associated with the mock dwarf galaxy.}
\label{fig:example_field}
\end{figure*}


\section{Search Algorithm}
\label{sec:algorithm}

Having prepared the \emph{Gaia} data for our analysis, we now provide an overview of the wavelet transform and its application in our search algorithm (Section \ref{wavelet_subsec}), the optimization of our algorithm for dwarf galaxy detection (Section \ref{algorithm_nitty_gritty}), and our procedure for measuring the significance of the overdensities returned by our search (Sections \ref{random_field_generation}--\ref{sec:candidate_significance}). 

\subsection{Wavelet and Algorithm Overview}
\label{wavelet_subsec}

The discrete wavelet transform (DWT) was introduced by \cite{Mallat}. It functions similarly to a standard Fourier transform, but replaces the delocalized sine function convolution kernel with locally oscillating basis functions. It does this by convolving a given field with individual wavelets from a given wavelet family. In general, a wavelet family is a set of functions that cover the whole of Fourier space. Wavelets in a given family have the same shape but different sizes and orientations. Thus, wavelet convolution can be thought of as a local Fourier transform that amplifies power at a given scale and orientation based on the properties of the wavelet used in the convolution. Since we expect dwarf galaxies to be approximately isotropic in projected position and proper motion space, this procedure can be simplified for our purposes by averaging over various orientations.

The wavelet transform allows for spatially localized frequency analysis. Specifically, the DWT consists of recursively applying discrete low-pass and high-pass wavelet filters intersected by an operation that downsamples the data vector by a factor of two. This analysis returns a full set of wavelet coefficients, $c$, which can be divided into scaling coefficients $c_J$ and wavelet coefficients $d_j$,
\begin{equation}
    c = c_J + \sum_{j=0}^{J-1} d_j\end{equation}
    
where $l$ indicates the maximum depth of the wavelet transform. The $c_J$ coefficients are the product of $l$ applications of the low-pass filter, while the $d_j$ coefficients are computed as $j$ convolutions with the low-pass filter followed by a convolution with the high-pass filter. Since the data vector is downsampled by a factor of two at each level $j$, the $d_0$ coefficients probe the smallest scales, whereas the $c_J$ coefficients probe the largest scales. By decomposing a given field at various scales, the wavelet transform allows for the selection of specific scales of interest; in particular, we can manually increase power on those scales by amplifying the corresponding coefficients before computing the inverse transform. Thus, by carefully selecting and amplifying scales expected to correspond to dwarf galaxies, we can increase the contrast between the dwarf and background signals to maximize our detection sensitivity.

The main steps of our wavelet dwarf galaxy search algorithm, which are described in detail in the following subsections, are as follows:
\begin{enumerate}
  \item Generate a 4D histogram of stellar positions and proper motions for a given \emph{Gaia} field.
  \item Transform the histogram using wavelet decomposition and amplify wavelet coefficients at specific scales.
  \item Perform the inverse wavelet transform and set all pixels below a given threshold to zero. 
  \item Identify remaining overdensities in the image and select all \emph{Gaia} stars within those overdensities.
\end{enumerate}
A visualization of these four steps for a field containing a mock dwarf galaxy is shown in Figure \ref{fig:analysis_pipeline}. 

\subsection{Optimization for Dwarf Detection}
\label{algorithm_nitty_gritty}

In order to use the wavelet transformation to aid in dwarf detection, we begin by generating
$96\times96\times96\times96$ 
4D histograms for a given stellar field.\footnote{We summarize all of the specific values we choose for our algorithm in Table \ref{tab:table6} in Appendix \ref{appendixa}.} An example stellar field before and after the parallax cut has been applied is shown in the top two panels of Figure \ref{fig:example_field}, which illustrates the field for the same mock dwarf galaxy as in Figure \ref{fig:analysis_pipeline}. The histogram size is chosen to be ($i$) fine enough so that we have approximately one star per bin in most fields, and ($ii$) divisible by $2^4$ to allow for a wavelet transform with depth $J = 4$. An example of this histogram projected into position and proper motion space is shown in the left column of Figure \ref{fig:analysis_pipeline}.

\begin{figure*}[htb!]
\centering
\includegraphics[width=\textwidth]{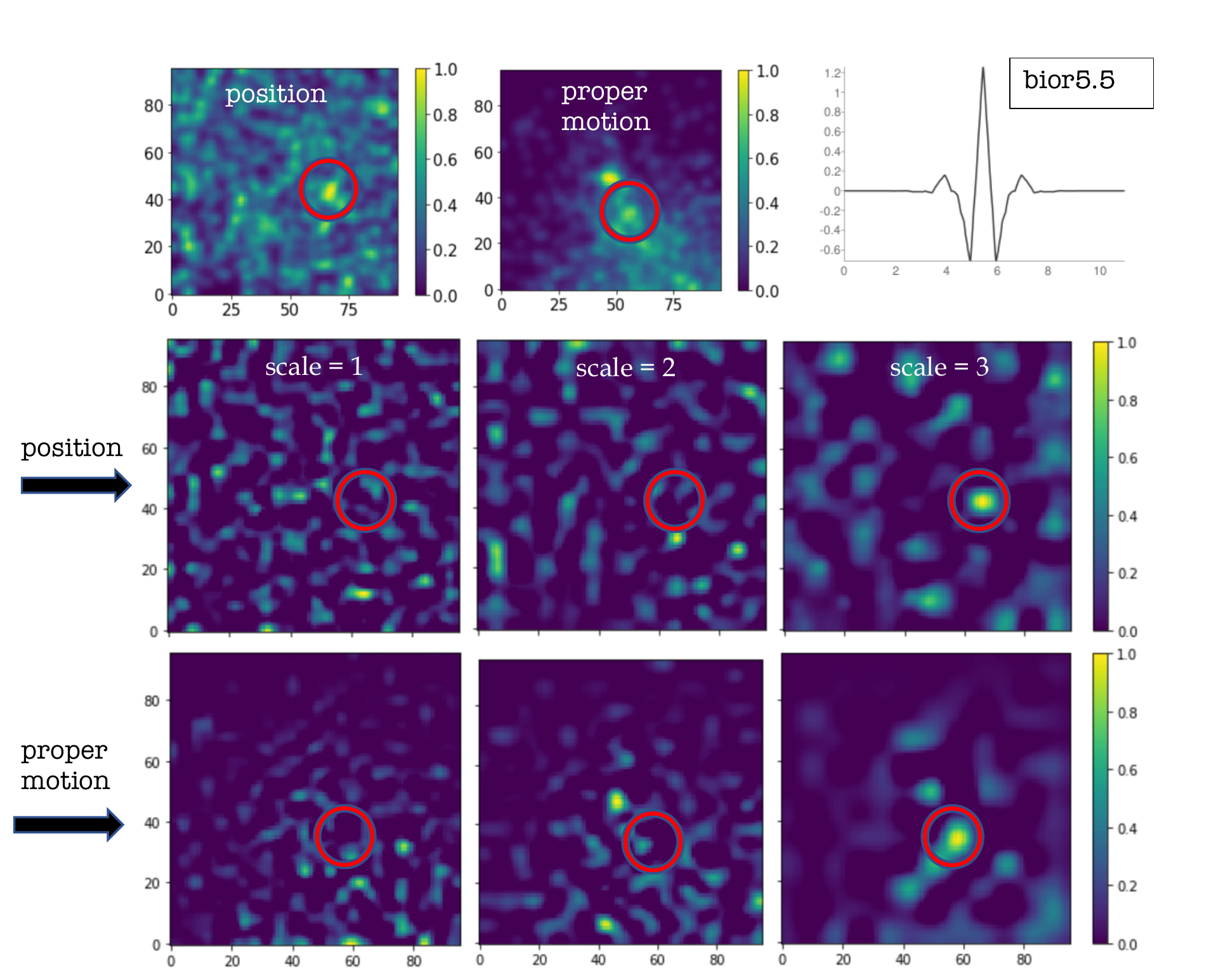}
\caption{Illustration of how our wavelet search algorithm amplifies signal at different scales in position and proper-motion space, using the Tucana II field as a reference. In all panels, the red circle highlights the region we expect to observe a phase-space overdensity based on the known position and proper motion of Tucana II. \emph{Upper row:} The histogram of \emph{Gaia} stars in position (\emph{left}) and proper motion (\emph{middle}) space before applying the wavelet transform. \emph{Top right:} The shape of the \texttt{bior5.5} wavelet, shown as an example to demonstrate the response of the transform as a function of scale \citep{Lee2019}. \emph{Lower row:} The position (\emph{middle row}) and proper motion (\emph{bottom row}) fields after applying the transform and amplifying only the indicated scale (i.e., Equation \ref{eq:djprime} for $j \in [1,2,3]$). Tucana II is highlighted particularly well at the $j = 3$ scale, as expected given its half-light radius of $r_{1/2} = 165\ \mathrm{pc}$ and distance of $D = 58\ \mathrm{kpc}$.}
\label{fig:scale_example}
\end{figure*}

Next, we compute a 4D DWT on the 4D position--proper motion histogram. This transform can be computed to various depths, with the maximum depth set by the number of times the downsampling operation can be performed. For this analysis we perform a $J=4$ transform on the 4D histogram. We amplify the transformed image at certain scales. In order to amplify the signal of a potential dwarf galaxy in the field, we boost the three largest-scale wavelet coefficients by a factor of 100: 
\begin{equation}
    d_j' = 100\times d_j,\ \mathrm{for}\ 1 \leq j < J.\label{eq:djprime}
\end{equation}
Multiple scales were chosen to improve the algorithm's ability to detect structure at a range of sizes, which helps account for the diversity in half-light radius and velocity dispersion of real dwarf galaxies \citep{Simon190105465}. The specific scales chosen here were selected via trial and error on known dwarf galaxies. An example for how this works on the dwarf galaxy Tucana II is shown in Figure \ref{fig:scale_example}. The figure shows the response to the amplification of each of the three scales used in this analysis individually. The signal of Tucana II is strikingly apparent at the $j=3$ scale, which is expected based on its large half-light radius ($r_{1/2} = 165\ \mathrm{pc}$) and nearby distance ($D = 58\ \mathrm{kpc}$). For known dwarfs with smaller sizes or at larger distances, the $j=1$ or $j=2$ scales show the strongest response. We do not amplify the smallest scale because it is below the angular size we expect to be sensitive to and was found to be very noisy.  Conversely, performing a deeper transform would have allowed us to amplify scales larger than $d_3$. However, this was found to wash out dwarf signals in the data. We note that the choice of which scales to amplify is degenerate with the fineness of our 4D binning. Here, we have chosen to maintain relatively fine binning to retain as much information as possible. We also experimented with amplifying fewer scales or different combinations of the chosen three scales; however, we found that amplifying all three scales recovered the highest percentage of known dwarf galaxies across the full range of dwarf parameter space (see Section \ref{sec:dwarf_analysis}).

We compute the DWT using wavelets from the bi-orthogonal wavelet family. Specifically we use the \texttt{bior5.5} wavelet implemented in \texttt{PyTorch} \citep{NEURIPS2019_9015}. We select this wavelet because ($\emph{i}$) its smooth distribution avoids unwanted amplification of small-scale structure, and ($\emph{ii}$) both its low- and high-pass filters have a unimodal distribution, corresponding to the single peaks in both position and proper motion space expected for dwarf galaxy stellar distributions. After amplification, we compute the inverse transform and co-add the resulting 4D histograms. Finally, we smooth the image using a Gaussian filter with $\sigma=3$; this scale is typically smaller than those amplified by the wavelet transform. An example of this procedure is shown in the top middle panel of Figure \ref{fig:analysis_pipeline}. It is visible by eye that the transform highlights structures at approximately the 10 pixel scale.


\subsection{Random Field Generation}
\label{random_field_generation}
In order to compare the significance of the peaks in the wavelet transform to a baseline, we generate a random field corresponding to each real field. The average density of stars in the original field is used to estimate a Poisson rate, which is averaged over 100 random samplings and multiplied by the total area of the field to get the number of stars for the random field. 


We fit a Gaussian mixture model (GMM) with $n=1-4$ components to the 2D proper motion distribution of stars in the field and sample from it in order to generate a random proper motion field. The number of components is chosen independently for each field by  minimizing the Bayesian information criterion (BIC). The number of samples is given by the total number of stars in the Poisson random field. We use a Gaussian mixture model because the stars in a given field are usually moving with a non-random bulk motion rather than being uniformly distributed across the projected phase space. We do not consider more than four components for the GMM to avoid overfitting the small-scale structure of the field. This process returns a random field that is visually similar to the real field, as shown in Figure \ref{fig:example_field} in both position and proper motion space. We treat the random field identically to the real field during the application of the wavelet transform.

\subsection{Candidate Identification}
\label{sec:candidate_significance}

We use the random fields described above to benchmark the significance of the candidates returned by our search. In particular, we standardize the histogram according to 
\begin{equation}
    p' = \frac{p - \mu_{p_{{\mathrm{rand}}}}}{n\sigma_{p_{{\mathrm{rand}}}}},
\label{eq:sigmaw}
\end{equation}
where $p$ is the pixel value of the transformed image and $\mu_p{_{\mathrm{rand}}}$ and $\sigma_p{_{\mathrm{rand}}}$ are the mean and standard deviation of the transformed random field, respectively. Here $n$ is a function of stellar density,
\begin{equation}
    n = -0.6\log_{10}({n_{\mathrm{stars}}}) + 4, 
\label{eq:correction}
\end{equation}
where $n_{\mathrm{stars}}$ is the total number of stars in the field. This function was fit based on our analysis of mock dwarf galaxies (Section \ref{sec:mock_analysis}) to convert the value of the transformed image to a more interpretable significance value. Specifically, the conversion was fit to transform the most significant pixel in the wavelet transform to approximate the Poisson significance of a given overdensity. We then set all pixels with $p' < 5$ to zero in order to select regions of high significance. An example of this thresholded image is shown in the third panel of Figure \ref{fig:analysis_pipeline}. Algorithm \ref{algorithm_wavelets} in Appendix \ref{appendixa} shows a schematic of the pipeline used to perform the wavelet analysis.\footnote{\href{https://github.com/edarragh/demeter}{https://github.com/edarragh/demeter}}

In order to identify distinct overdensities in position and proper motion space that have been amplified by the wavelet transform, we begin by cutting out the outer 1/8 of the image to avoid edge effects introduced by the transform. We then label connected non-zero regions in 4D space. In particular, we consider two pixels to be connected if they are adjacent in 4D (each pixel has 32 neighbors). For each connected region, we fit a single component GMM to the 2D distributions in position and in proper motion space. We then define the position and the radius of the candidate as
\begin{align}
    & x, y = \mu_{GMM} & \\ 
    & r_x, r_y = 4\sqrt{\sigma_{GMM}^2} & 
\label{eq: radii}
\end{align}
where $x$, $y$ are the candidate center in position (proper motion) and $r_x$, $r_y$ are the candidate radii in position (proper motion) from the 2D GMM fit in position (proper motion) space. These radii are defined to be twice the naive estimate to avoid selection against spatially extended UFDGs by ensuring that we include stars in the outer regions of objects, which might have been removed during the thresholding procedure. We then associate a wavelet significance with each patch as the value of the most significant pixel in the original 4D region. We emphasize that this ``significance'' has been normalized to approximate the Poisson significance of the overdensity in position space (see Section \ref{sec:mock_analysis}). 

Lastly, we select clusters of stars that fall within the returned spheres in position and proper motion space, combining clusters that share one or more stars. We only consider clusters that return $\geq 5$ stars to remove spurious clusters, especially in underdense fields. While this cut has the potential to remove real faint objects, these systems will hopefully be found in runs on future data releases with more complete stellar catalogues and more precise astrometry. An example of the stars identified as belonging to the mock dwarf galaxy considered in this example can be seen in grey in Figure \ref{fig:analysis_pipeline}. In this example, no other candidates have been identified in the field. The algorithm recovers member stars with a completeness of $63\%$ and a purity of $60\%$. We can see in the bottom panel of Figure \ref{fig:example_field} that the algorithm does an excellent job of recovering stars in the core of the mock dwarf. However, it does miss several of the stars lying at significantly larger radii. As expected, the contamination from unassociated stars in the field comes from stars that happen to lie in similar regions in projected position and proper motion, highlighting the need for precise distance measurements to confirm candidates returned by our search. Nevertheless, the main signal is recovered at high significance (wavelet significance $\sigma_w = 11.0$ and Poisson significance $\sigma_p = 8.1$; see Section \ref{sec:mock_analysis}) and at relatively high purity. For a technical description of the process of candidate identification and star selection, see Algorithm \ref{algorithm_stars} in Appendix \ref{appendixa}.

\section{Significance Calibration}
\label{sec:analysis}

\subsection{Mock Dwarf Galaxy Search}
\label{sec:mock_analysis}

Mock dwarf galaxies from \cite{Antoja150704353} were used to calibrate the significances of the overdensities returned by our analysis. The mock galaxies were generated using a Plummer sphere with an isotropic velocity dispersion. The mock galaxies span a range of parameters in luminosity, half-light radius, stellar mass, and distance (Table \ref{tab:mock_properties}). The stellar populations were modelled as a single burst of star formation with an age of 12 Gyr and metallicity of $Z=0.0001$ assuming a Chabrier IMF with stars sampled to represent realistic \emph{Gaia} DR2 magnitude limits and errors in stellar parameters. In total, 6000 mock galaxies were used in the analysis with parameters drawn at random from the distributions in Table \ref{tab:mock_properties}. As described by \cite{Antoja150704353}, the sampling over these parameters is not uniform as a function of distance -- both to conservatively approximate the detectability of known dwarf galaxies and to remove regions of parameter space where we do not reasonably expect to find undetected systems (i.e. very nearby bright systems). In order to understand the sensitivity of our algorithm as a function of stellar density the mock dwarfs were inserted into 14 different fields in the \emph{Gaia} data. These fields were located at two Galactic longitudes $(\ell=90^{\circ},\ 350^{\circ})$ and at Galactic latitudes at $10$ or $20$ degree increments $(b \in [10^{\circ}, 20^{\circ}, 30^{\circ}, 40^{\circ}, 50^{\circ}, 70^{\circ}, 90^{\circ}])$, with denser sampling at low latitudes due to the steep gradient in average stellar density versus Galactic latitude. The number of stars in these fields ranged from approximately 1500 stars at $b=90^{\circ}$ to approximately 100,000 stars at $b=10^{\circ}$. In an attempt to control for any possible bias introduced by the \emph{Gaia} field chosen, 3000 mocks were run in the initial $\ell=90^{\circ}$ field and 3000 in the $\ell=350^{\circ}$ field at each latitude. Some of the $\ell=350^{\circ}$ fields had significantly more stars than the original field at $\ell=90^{\circ}$. For these fields, the stars were randomly downsampled to be within $10\%$ of the number of stars in the original field.

\begin{deluxetable}{l c c }
\centering
\tabletypesize{\footnotesize}
\tablecaption{\label{tab:mock_properties} Range of parameters for mock dwarf galaxies used to estimate our search sensitivity (Section \ref{sec:mock_analysis})}
\tablehead{\colhead{Dwarf Parameter} & \colhead{Range of Values} & \colhead{Unit} }
\startdata
Luminosity &  $[86,7.5 \times 10^6]$ & $\mathrm{L}_{\mathrm{\odot}}$ \\
Half-light radius & $[5,4000]$ & $\mathrm{pc}$ \\
Velocity dispersion & $[1,500]$ & $\mathrm{km\ s}^{-1}$ \\
Heliocentric distance  & $[10,250]$ & $\mathrm{kpc}$ \\
Mass-to-light ratio & $[0.04,3.9 \times 10^7]$ & $\mathrm{M}_{\mathrm{\odot}}\ \mathrm{L}_{\mathrm{\odot}}^{-1}$ \\
Stellar mass & $[10,1.6 \times 10^6]$ & $\mathrm{M}_{\mathrm{\odot}}$ \\
Number of observed stars & $[10,10^3]$ & \ldots \\
\enddata
{\footnotesize \tablecomments{These parameters are not uniformly sampled as a function of heliocentric distance (see \citealt{Antoja150704353}).}}
\end{deluxetable}


    
    
In order to understand how the peak value returned by the wavelet transform compares to a standard Poisson significance, we began by running the algorithm over the mock galaxies with no correction to the transformed image ($n = 1$; Equation \ref{eq:sigmaw}) and a flat threshold at $p' = 20$. For each returned overdensity we calculated a Poisson significance according to
\begin{equation}
    \sigma_{p} = \frac{N_{obs} - \langle N_{obs}\rangle}{\sqrt{\langle N_{obs}\rangle}},
\end{equation}
where $N_{obs}$ is the number of stars in the overdensity and $\langle N_{obs}\rangle$ is the number of stars we would expect within the area of the overdensity assuming a purely Poisson random distribution. We estimate $\langle N_{obs}\rangle$ by filtering the stars in the full field based on the extent of the overdensity in proper motion space (Equation \ref{eq: radii}) and then estimating the Poisson rate analogously to how it was done when generating the random field (Section \ref{random_field_generation}). This Poisson rate was then used to determine the expected number of stars given the area of the candidate. 

\begin{figure*}[htb!]
\centering
\includegraphics[width=.98\columnwidth]{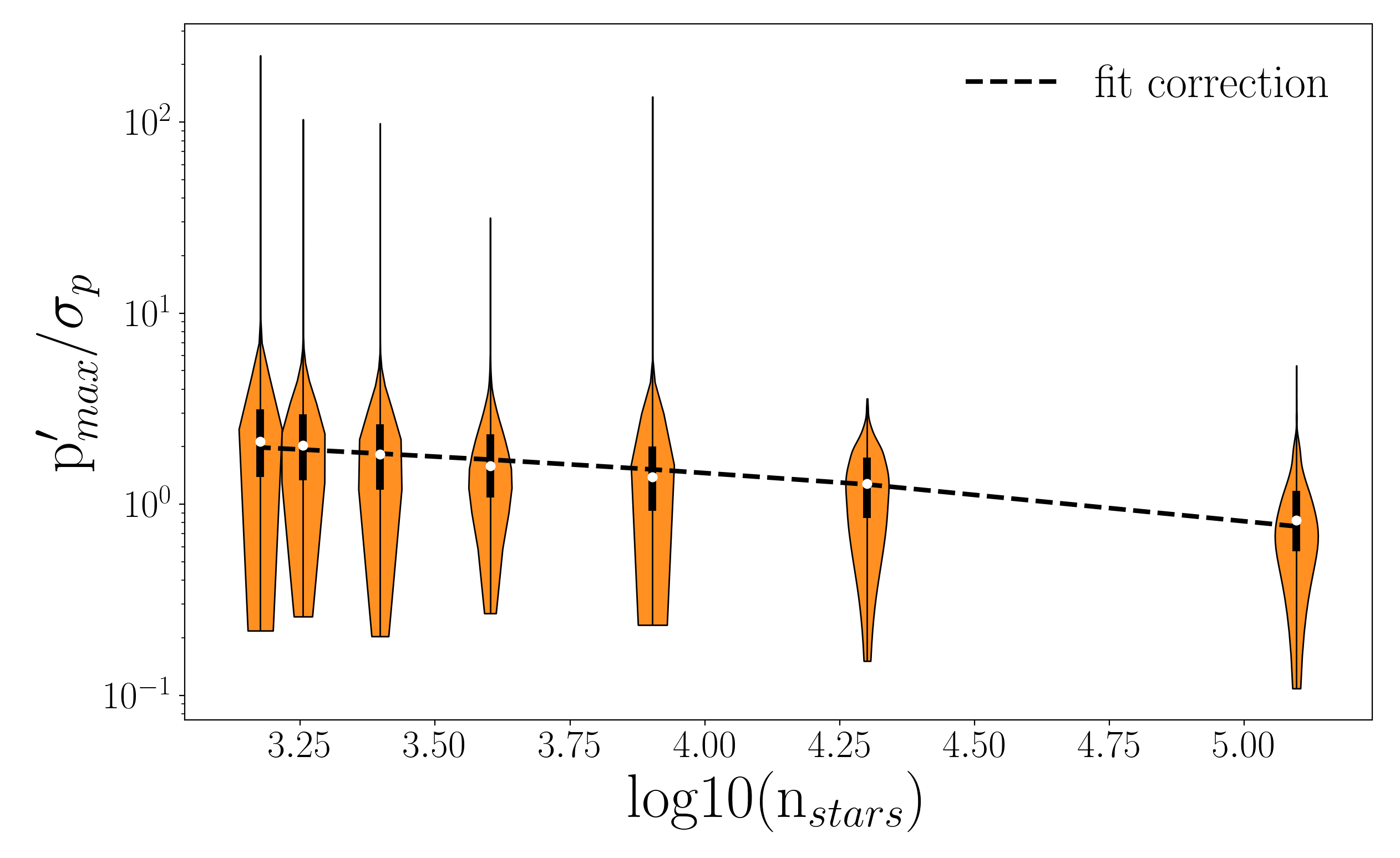}
\includegraphics[width=1.02\columnwidth]{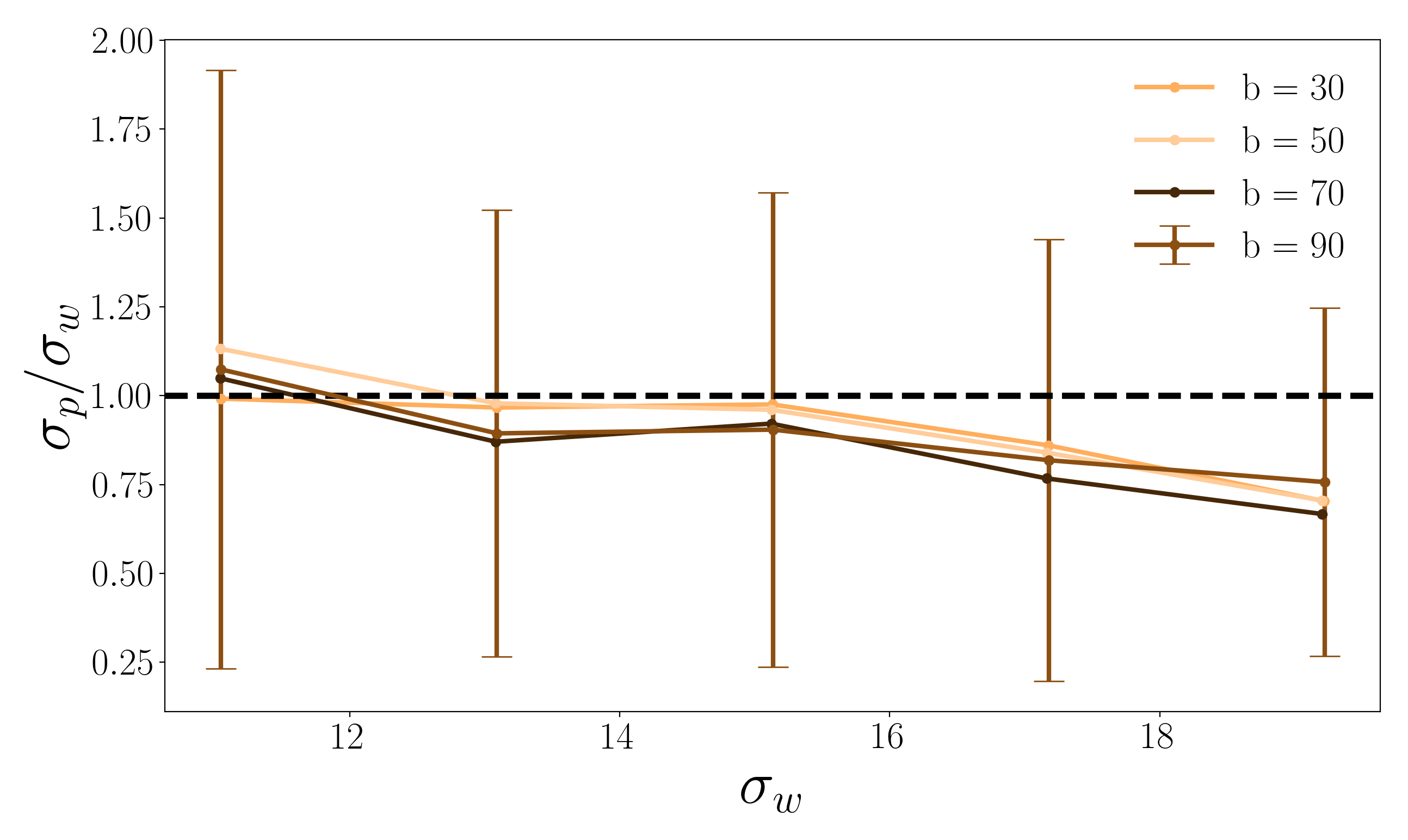}
\caption{\emph{Left:} The ratio of the peak overdensity height returned by our wavelet search algorithm, $p’_{\mathrm{max}}$, to the Poisson significance in position space, $\sigma_p$, as a function of local stellar density $n_{\mathrm{stars}}$ from our search on mock dwarf galaxies (Section \ref{sec:mock_analysis}). The orange violin plots represent the range of values for all recovered mock galaxies at the individual stellar densities probed, while the dashed line represent the fit correction across all stellar densities. \emph{Right}: The mean and $68\%$ confidence interval of the ratio of $\sigma_p$ to the wavelet significance corrected to approximately correspond to a Poisson significance, $\sigma_w$, as a function of $\sigma_w$. Colors indicate bins of Galactic latitude for all mock galaxies detected with uncorrected wavelet significances $p'_{max} < 50$. Confidence intervals are only shown for $b = 90^{\circ}$; however, the sizes of these confidence intervals are consistent across all latitudes $10^{\circ}<b<90^{\circ}$. The ratio of wavelet versus Poisson significance is consistent with a constant value of unity for all $\sigma_w$ and at all latitudes.}
\label{fig:correction}
\end{figure*}

From this analysis, we were able to estimate the necessary correction to make the wavelet peak height, $p'_{max}$, approximately equal to the Poisson significance, $\sigma_p$, as a function of stellar density. This was done by fitting a constant correction to the ratio $\sigma_p/p'_{max}$ at each stellar density for $p'_{max} < 50$. Restricting to low-signal objects allows us to focus on the regime where we expect to be sensitive to undiscovered dwarf galaxies relative to previous searches (see Section \ref{sec:mock_analysis}). In doing so, we exclude the high-signal tail dominated by relatively bright mock dwarfs, which should be detected in previous surveys except at very low latitudes. 
The mocks used in the fit are shown in Figure \ref{fig:correction}, where the additional cut of $0 < \sigma_p/p'_{max}$ was implemented to remove objects where $\sigma_p$ is poorly measured.

We fit to the average correction to $\sigma_p/p'_{max}$, as a function of $\log_{10}(n_{\mathrm{stars}})$. A linear fit was found to well approximate the change in offset as a function of the number of stars in the field (Figure \ref{fig:correction}). The error bars on this fit are very large and even consistent with a constant correction. However, based on our analysis on known dwarf galaxies (Section \ref{sec:dwarf_analysis}), a linear correction with stellar density increases the sensitivity of our search uniformly across all latitudes.

After determining the correction, the analysis was re-run on the full library of 6000 mocks in all 14 fields described above. The correction was applied before the thresholding step of the wavelet algorithm. All pixels with a corrected value $p' < 5$ were set to zero. The accuracy of the correction was checked by plotting the corrected ``wavelet significance'' of a given candidate, $\sigma_w$, versus $\sigma_p/\sigma_w$ as a function of $\sigma_w$ for all 14 fields and for dwarfs with an uncorrected peak height of $p'_{max} < 50$. The ``wavelet significance'' is given by:
\begin{equation}
   \sigma_w = p'_{max},
\end{equation}
where $p'_{max}$ is the maximum value of $p'$ in the candidate region after normalizing the transformed histogram according to Equation \ref{eq:sigmaw}. In Figure \ref{fig:correction} we see that there is a slight trend toward overcorrection at higher values of $\sigma_w$. However, all of the corrected means are consistent with unity at the $68\%$ confidence level, and we do not observe significant differences in the correction as a function of Galactic latitude. 

\subsection{Search Sensitivity on Mock Data}
\label{sec:sensitivity}

\begin{figure*}[htb!]
\centering
\includegraphics[width=\textwidth]{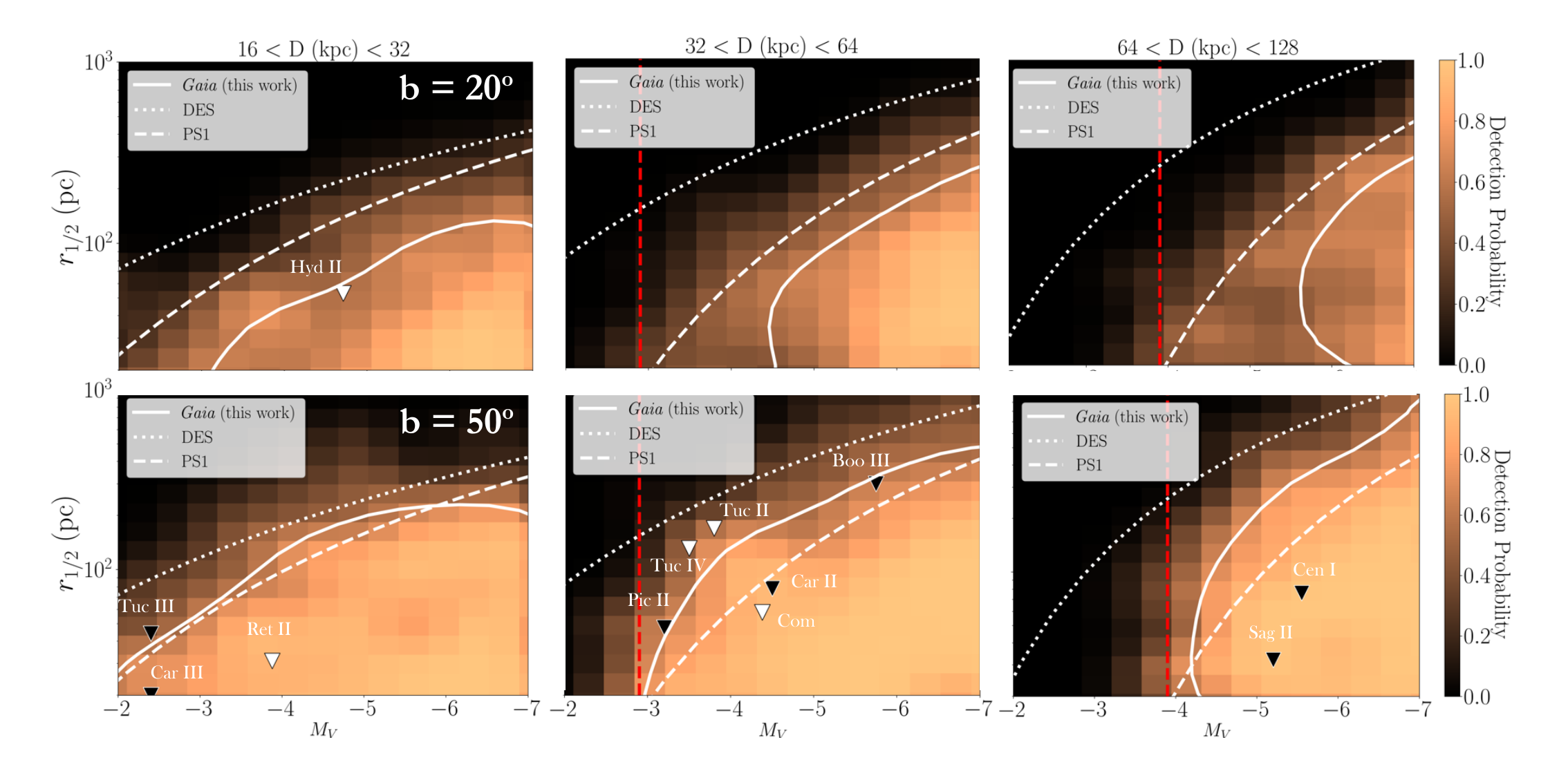}

\caption{The sensitivity of our phase-space search for dwarf galaxies in \emph{Gaia} DR2 data compared to the sensitivity of photometric searches in DES (dotted) and PS1 (dashed) data for $b = 20^{\circ}$ (\emph{top row}) and $b = 50^{\circ}$ (\emph{bottom row}), in three bins of heliocentric distance: $16\ \mathrm{kpc}<D<32\ \mathrm{kpc}$ (\emph{left column}), $32\ \mathrm{kpc}<D<64\ \mathrm{kpc}$ (\emph{middle column}), and $64\ \mathrm{kpc}<D<128\ \mathrm{kpc}$ (\emph{right column}). The \emph{Gaia} sensitivity color map is calculated from our analysis on mock dwarf galaxies and smoothed using a $1\sigma$ Gaussian kernel. The dashed red line marks the faintest mock dwarf galaxy included in our analysis. The solid white line represents the estimated $50\%$ sensitivity contour for our search. The white triangles represent known dwarfs detected by our search, while the black triangles correspond to known dwarfs that are not recovered by our analysis.} 
\label{fig:sensivity}
\end{figure*}

Next, we used the mocks to estimate the sensitivity of the algorithm as a function of stellar density, magnitude, half-light radius, and distance. This was done separately for the different Galactic latitudes. An example of our search sensitivity at $b=50^{\circ}$ can be seen in the top panel of Figure \ref{fig:sensivity}. Our search is strictly less sensitive than the search in DES Y3A2 photometric data \citep{Drlica-Wagner2020}. However, we expect to recover dwarfs more efficiently than the PS1 search at distances between $16\ \mathrm{kpc} < D < 128\ \mathrm{kpc}$. Specifically, we expect to be sensitive to undetected objects larger than those corresponding to the $50\%$ contour of the PS1 sensitivity function between $-7<M_V<-3$ for $32\ \mathrm{kpc} < D < 64\ \mathrm{kpc}$ and for objects brighter than $M_V = -4$ for $64\ \mathrm{kpc} < D < 128\ \mathrm{kpc}$. The red dashed lines in Figure \ref{fig:sensivity} mark the faintest mock objects analyzed, so it is possible the algorithm could be sensitive at slightly fainter magnitudes, but we find a sharp drop off in detection efficiency below this limit. 
At $D > 128\ \mathrm{kpc}$, we observe a sharp drop-off in sensitivity relative to both the DES and PS1 searches. While we have only chosen to show $b = 50^{\circ}$, this trend is found to hold true across all latitudes with $\lvert b\rvert > 30^{\circ}$. 

For $\lvert b\rvert \leq 30^{\circ}$, the algorithm's sensitivity degrades rapidly (detection sensitivities for $b=20^{\circ}$ are shown in the bottom row of Figure \ref{fig:sensivity}). However, because most of this region is excluded in photometric searches due to dust contamination \citep{Drlica-Wagner2020}, we nevertheless include candidates down to $\lvert b\rvert > 20^{\circ}$. Below $b = 20^{\circ}$, our search has limited detection sensitivity for objects at $D <64\ \mathrm{kpc}$. By visually examining these low-latitude candidates, we found that many are contaminated with foreground stars due to extremely high stellar densities. This is a common challenge for past dwarf searches at low latitudes, and may actually represent an interesting avenue for future work given the unique ability of our algorithm to leverage both position and proper motion information. However, we have chosen to ignore these candidates for the current analysis. Thus, we restrict the remainder of this analysis to regions outside of the DES footprint and at $\lvert b\rvert > 20^{\circ}$.

Based on the sensitivity predictions above, we estimate the number of undiscovered dwarfs expected to be recovered by this analysis using the galaxy--halo connection model from \cite{Nadler191203303}, which is fit to the DES and PS1 satellite populations based on the observational selection functions from \cite{Drlica-Wagner2020}. In particular, using the estimated search sensitivity shown in Figure \ref{fig:sensivity}, we use this model to predict the number of undiscovered satellites in the PS1 footprint in regions of ($M_V$, $r_{1/2}$, $D$) parameter space where our search is more sensitive than PS1. We find that our search is expected to detect $5\pm 2$ new satellite galaxies, where the error bar indicates both galaxy--halo connection uncertainties (which are minor relative to the statistical uncertainty for the dwarf galaxies of interest) and Poisson noise. Four are expected to be found in PS1, excluding regions of the PS1 footprint covered by DES, and one other is expected to be discovered by our search technique in regions of the $\lvert b\rvert >20^{\circ}$ sky that are not covered by DES or PS1 (see Figure \ref{fig:dwarf_position}). This remaining area has largely been observed by the DELVE\footnote{\href{https://delve-survey.github.io/}{https://delve-survey.github.io/}} survey, which is deeper than PS1; satellite searches in DELVE data are ongoing and have already revealed several promising candidates  \citep[e.g.][]{Mau191203301,Cerny200908550}.

\subsection{Search Sensitivity on Known Milky Way Satellite Galaxies}
\label{sec:dwarf_analysis}

\begin{figure*}[htb!]
\centering
\includegraphics[width=\textwidth]{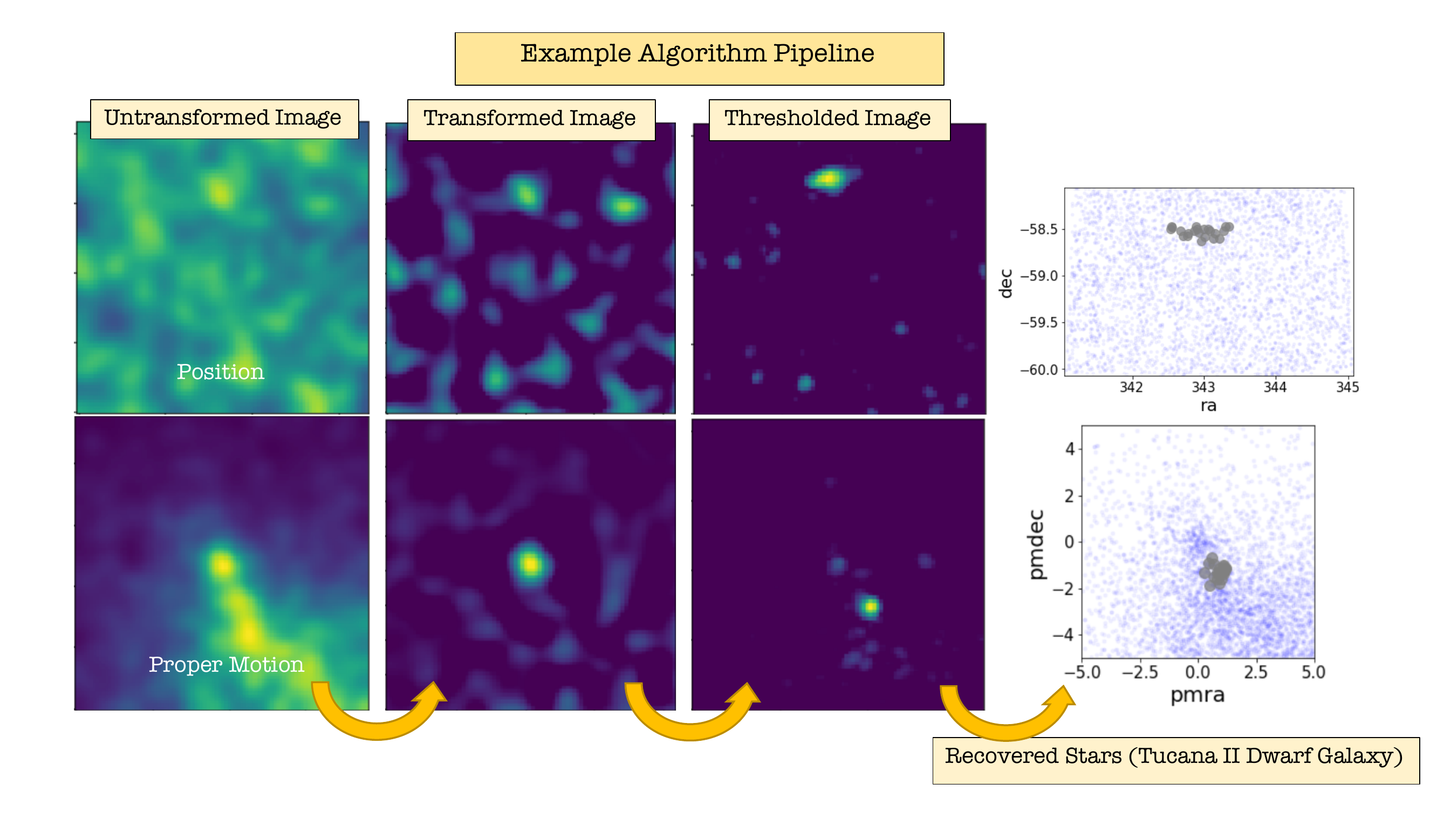}
\caption{Example of our dwarf galaxy search algorithm pipeline for the \emph{Gaia} field containing Tucana II. \emph{First column:} The untransformed histogram of \emph{Gaia} stars smoothed using a Gaussian filter and projected into position space (\emph{top}) and proper motion space (\emph{bottom}). \emph{Second column:} The transformed histogram of \emph{Gaia} stars after normalizing the image to a corresponding random field with the same Poisson rate according to Equation \ref{eq:sigmaw}. \emph{Third column:} The transformed histogram after thresholding on normalized overdensities with Poisson significances greater than $\sim 5\sigma$ (see Section \ref{sec:candidate_significance}). The left three histograms have all been normalized to take values between [0,1]. \emph{Fourth column:} The position and proper motion of the stars returned by our algorithm. The grey points show the stars returned by our algorithm for this field. Tucana II is returned with a Poisson-corrected wavelet significance of $\sigma_w$ = 11.1}
\label{fig:analysis_pipeline_tuc}
\end{figure*}

The sensitivity of the algorithm was also tested by running on fields containing known dwarfs. We focus on the satellites listed in \cite{Drlica-Wagner2020} excluding the very bright Large and Small Magellanic Clouds and Sagittarius (although we do include Fornax and Sculptor). This leaves us with 57 known dwarfs that we include in this analysis. An example of the algorithm run at the position of the known dwarf galaxy Tucana II ($M_V = -3.8$, $D = 58\ \mathrm{kpc}$, $r_{1/2} = 165\ \mathrm{pc}$) can be seen in Figure \ref{fig:analysis_pipeline_tuc}. We highlight the algorithm's performance on this dwarf because it is nearby, relatively faint, and has a relatively large half-light radius given its luminosity, making it a good fit for the type of object we expect to be most sensitive to in this analysis. The stars determined to belong to the dwarf are shown in the bottom panel. The algorithm appears to do a good job of recovering the signal expected to lie at (343.0, -58.6) degrees in position and (0.9, -1.2)  mas yr$^{-1}$ in proper motion. It is worth noting that the position in proper motion space does not correspond to the most visually prominent 2D overdensity seen in the top panel of the proper motion plane (Figure \ref{fig:analysis_pipeline_tuc}). This highlights the power of performing this search on the full 4D distribution as well as our algorithm's sensitivity to detect tightly clustered objects in proper motion space.
 
\begin{figure*}[htb!]
\centering
\includegraphics[width=0.9\textwidth]{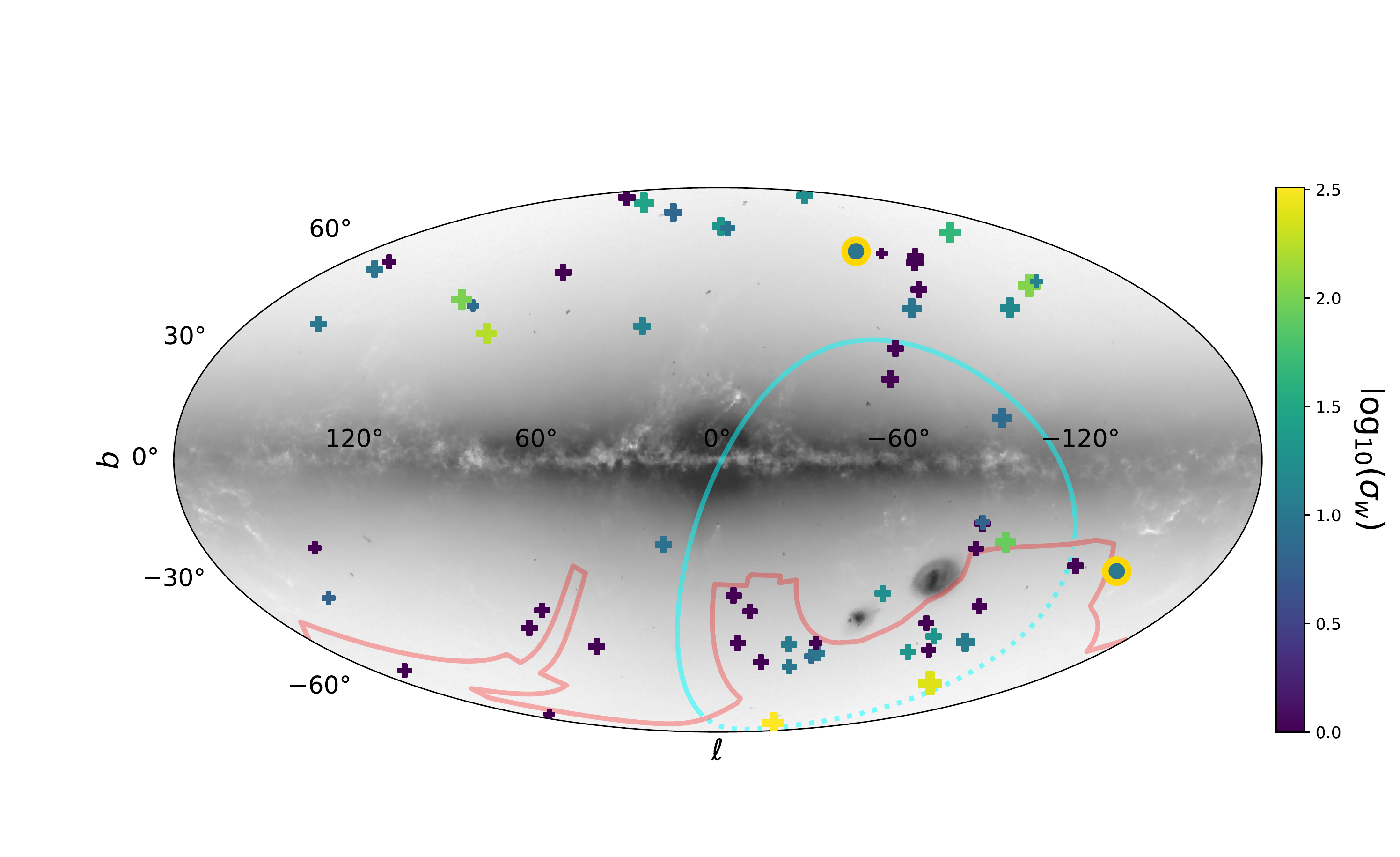}
\caption{Positions of known dwarf galaxies plotted over the background stellar density taken from the \emph{Gaia} DR2 data. Dwarfs are colored by their significance in our wavelet search on \emph{Gaia} data, corrected to approximately correspond to a Poisson significance, $\log_{10}(\sigma_w)$. Larger marker sizes indicate dwarfs with brighter absolute magnitudes. The \emph{plusses} correspond to known dwarfs, while the \emph{circles} outlined in gold represent the two candidates from our ``gold standard'' list that have an associated \emph{Gaia}-identified RR Lyrae star. The red line indicates the boundary of the DES footprint, the cyan line bounds the region covered by PS1, the region above the red curve and inside the cyan circle is covered by neither survey.}
\label{fig:dwarf_position}
\end{figure*}

The positions of known dwarfs are plotted in Figure \ref{fig:dwarf_position} where the recovered dwarfs are colored by $\log_{10}(\sigma_w)$. After running on the full catalogue of 57 objects, we recover 23 at high significance ($\sigma_p$ > 8.0 and $\sigma_w$ > 9.0) and another 8 are at low significance ($\sigma_p$ < 8.0 or 5.0 < $\sigma_w$ < 9.0). The remaining 26 dwarfs are not recovered by the algorithm at all. Most of these are either very small (15 have $r_{1/2} < 50\ \mathrm{pc}$), very faint (13 have $M_V > -3.5$), or distant (11 have $D > 132\ \mathrm{ kpc}$). This can be seen in the left panel of Figure \ref{fig:dwarf_properties}, which shows recovered dwarfs as a function of $M_V, D,$ and $r_{1/2}$. We therefore choose a cutoff of $\sigma_w > 9.0$ for our search in order to maximize purity and completeness given our tests on mock data and known systems. As shown in the left panel of Figure \ref{fig:dwarf_properties}, the algorithm performs best for real dwarfs that are bright ($M_V < -3.5$), nearby ($D < 200\ \mathrm{kpc}$) and have relatively large half-light radii ($100\ \mathrm{pc} < r_{1/2} < 1000\ \mathrm{pc}$). Out of the 15 known dwarfs in this range, the algorithm recovers 12/15 at high significance. This recovery rate is in good agreement with the predictions from the mock analysis. The performance in this region of parameter space also suggests that the algorithm can be used for globular cluster detection (also see \citealt{Huang200514014}); this is an interesting area for future work.

The completeness curve for the known dwarfs as a function of stellar density can be seen in the right panel of Figure \ref{fig:dwarf_properties}. Here, we have only included real dwarfs that lie within the distance-dependent range of magnitudes and half-light radii sampled by the mocks. This cut removes about half (27) of the known dwarfs. Despite high scatter due to low statistics, our search completeness for the remaining known dwarfs is reasonably consistent with our completeness on mock data.  

We also estimate the purity of our search. For fields with a dwarf object, the purity is very consistent between the mock and dwarf fields, and indicates purity greater than $80\%$ for all latitudes with $\lvert b\rvert >30^{\circ}$. However, it is worth noting that, due to the construction of our wavelet algorithm, the presence of high-significance candidates in a given field lowers the significance of other candidates in the same field. Thus, while the purity is high for fields containing a known object, it will likely be lower across the full sky. 

Finally, we compare the average proper motion values of recovered stars to published values for the Milky Way dwarfs from \cite{Fritz180500908}. We find that on average we recover the $\mu_{\alpha}$ value to within $-0.05 \pm 0.15$ mas yr$^{-1}$ and the $\mu_{\delta}$ value to within $0.05 \pm 0.32$ mas yr$^{-1}$ for the dwarfs recovered at high significance. Given that the \emph{Gaia} DR2 error bars on these measurements are sometimes over $100\%$, this agreement is highly encouraging.

Lastly, we also check the algorithm's performance on recovering known dwarfs in the recently released \textit{Gaia} EDR3 data. Since the algorithm was optimized to run on DR2 data and tested on mocks with DR2-like errors, we do not expect our sensitivity estimates or performance to translate directly to EDR3. Nevertheless, we find a higher recovery rate of known dwarfs across almost all stellar densities in the EDR3 data (Figure \ref{fig:dwarf_properties}). This indicates that the improved astrometric measurements and catalog completeness of future \textit{Gaia} releases and proposed astrometric surveys including \emph{Theia} \citep{Theia} will significantly improve the power and sensitivity of this approach. Furthermore, we plan to perform a tailored sensitivity analysis and search on EDR3 data in future work.

\section{Results}
\label{sec:full_sky_analysis}

Once we were satisfied with the performance of the algorithm, we extended the analysis to a run over the full sky. This analysis returned around 30,000 candidates. In order to reduce this to a manageable number we apply the following cuts. First, we remove known objects using the catalog compiled by \cite{Drlica-Wagner2020}. This catalog includes dwarfs, globular clusters, and open clusters. Next, we remove objects that fall within the DES footprint or at $\lvert b\rvert \leq 20^{\circ}$ (see Section \ref{sec:mock_analysis}). Lastly, we restrict to candidates with $\sigma_w > 9.0$ and $\sigma_p > 8.0$ (see Section \ref{sec:mock_analysis} and \ref{sec:dwarf_analysis}). After these steps, we were left with a sample of around 1300 unique candidates of interest. A significant number of these candidates are found to overlap with the Sagittarius Stream. We remove these candidates using the full sky proper motion distribution of the Sagittarius Stream using data from \citep{Antoja_2020}. In particular, we bin stream member stars using an NSIDE=8 HEALPix map. For each candidate from our search, we compare the total proper motion ($\mu_{tot} = \sqrt{\mu_{\alpha}^2 + \mu_{\delta}^2}$ mas yr$^{-1}$) to the average total proper motion of member stars in that pixel and remove candidates with total proper motion within $\pm 1.5$ mas yr$^{-1}$ of the stream. We also remove candidates at $\delta_{2000} < -60^{\circ}$, which are highly contaminated by the outskirts of the LMC and SMC. After applying these cuts we are left with approximately 830 high significance ``clean'' candidates.

The highest significance candidate in the ``clean'' list has a significance of $\sigma_w=14.7$. Out of all of the candidates, 75 are at $\lvert b\rvert \leq 30^{\circ}$. There are also 26 candidates that fall outside of the PS1 footprint. These candidates are particularly interesting because they fall outside of the region \cite{Drlica-Wagner2020} included in the DES and PS1 photometric search. Additionally, 11 of these candidates have an associated RR Lyrae star identified from the \emph{Gaia} RR Lyrae catalog. Most known dwarfs have been found to have associated RR Lyrae \citep{Vivas_2020}, so we consider these candidates to be especially interesting. However, since the \emph{Gaia} RR Lyrae catalogue is only expected to be 60\% complete due to a limited number of observation epochs, we still intend to follow up candidates without an associated RR Lyrae. 

\begin{figure*}[htb!]
\centering
\includegraphics[width=1.025\columnwidth]{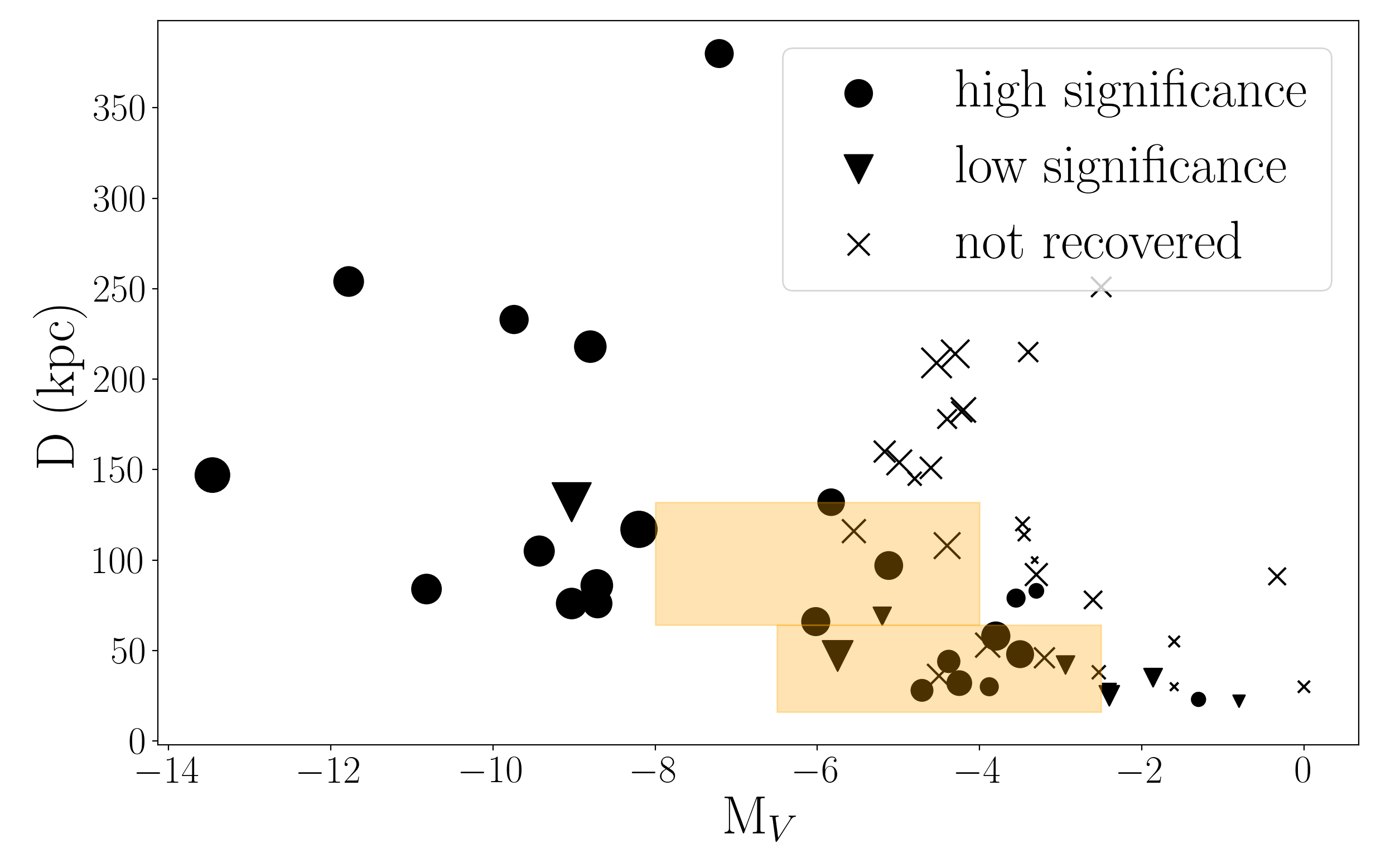}
\includegraphics[width=.9975\columnwidth]{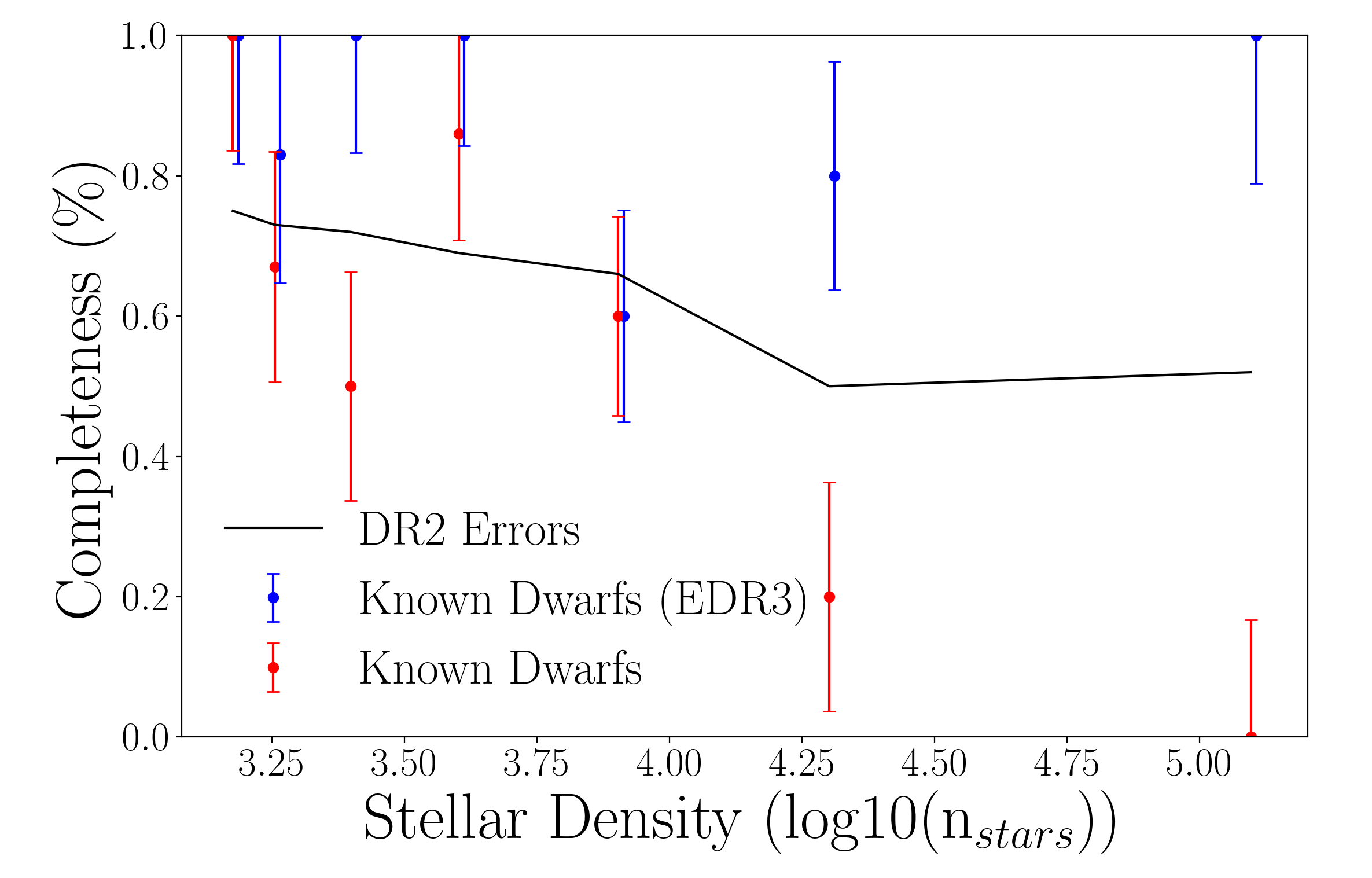}
\caption{\emph{Left:} Absolute magnitude vs distance of known dwarfs, with marker size indicating half-light radius. The different shapes correspond to dwarfs returned at high significance (\emph{circles}), low significance (\emph{triangles}), and those not recovered by the algorithm (\emph{crosses}; see Section \ref{sec:mock_analysis}). The orange boxes represents the region of parameter space in which we expect our search to be sensitive to undiscovered dwarf galaxies. \emph{Right:} Completeness of our phase-space search in \emph{Gaia} data for mock dwarf galaxies as a function of Galactic latitude for mock galaxies, and from our search for known dwarf galaxies in DR2 (red points) and EDR3 (blue points). Here we have removed known dwarfs that lie outside of the parameter space probed by the mocks. The errors on the dwarfs are estimated using jackknife resampling including the two adjacent bins.}
\label{fig:dwarf_properties}
\end{figure*}

In order to further refine this list, we cross-match the candidates from our DR2 search with candidates returned by applying the same algorithm to EDR3 data. As mentioned in Section \ref{sec:dwarf_analysis}, rigorous analyses of the EDR3 data is ongoing and requires tailored mock catalogs for validation. However, given the increased sensitivity to known dwarfs of the initial run (Figure \ref{fig:dwarf_properties}), we expect the candidates that are returned by both searches to have a higher likelihood of being real objects. For EDR3 we slightly increase our significance cutoff to  $\sigma_w > 10.0$ based on analysis of the known dwarfs. We match 232 objects in the ``clean'' list to candidates in the EDR3 list within one degree. The resulting ``gold standard'' list contains 12 candidates at $\lvert b\rvert \leq 30^{\circ}$. Additionally, 6 candidates lie outside of the PS1 footprint and 2 contain a known \emph{Gaia}-identified RR Lyrae.

\begin{figure}[t]
\centering
\includegraphics[width=\columnwidth]{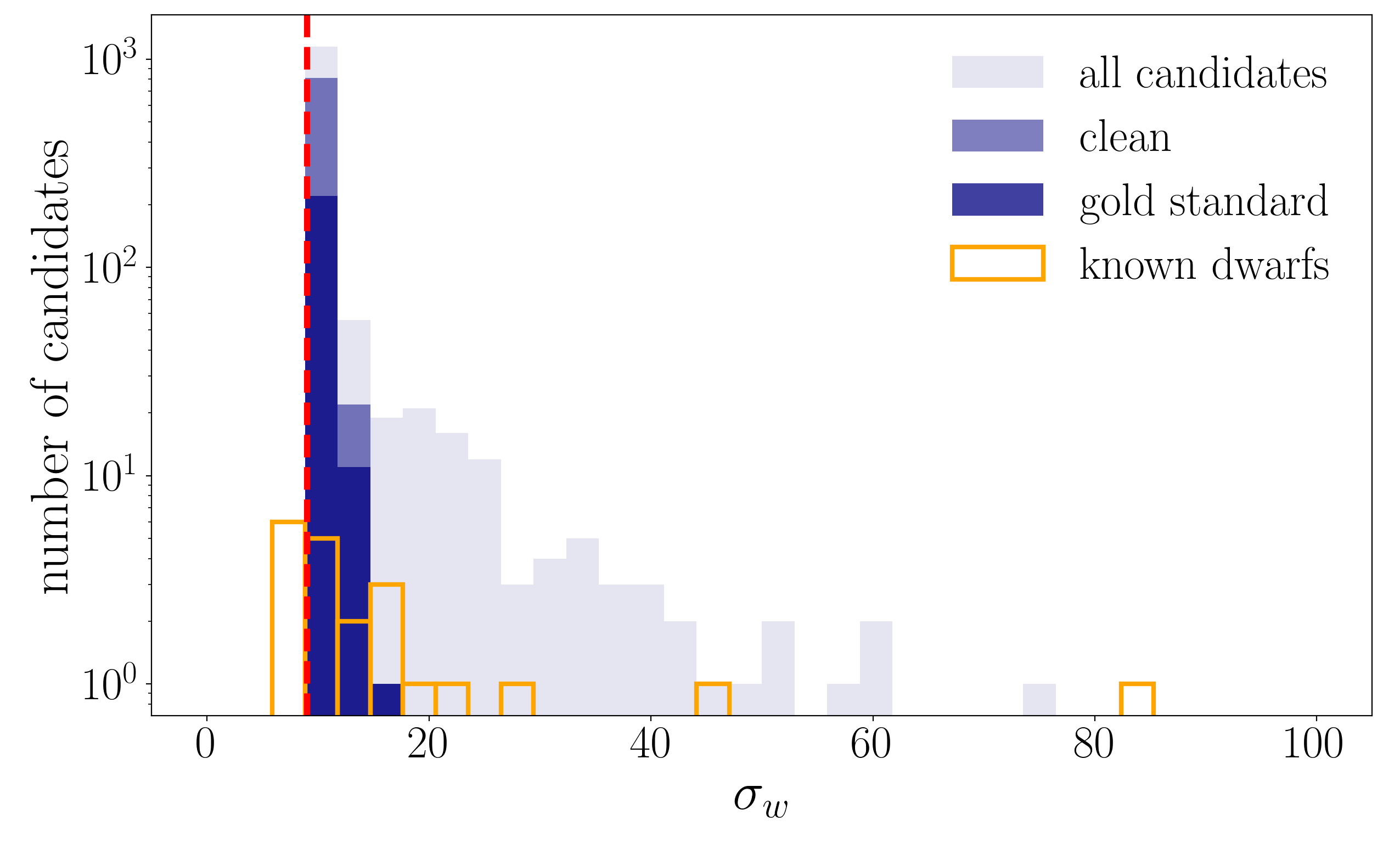}
\caption{The number of candidates returned by our search on \emph{Gaia} DR2 data versus the corrected wavelet significance $\sigma_w$. The faintest histogram shows all candidates after removing known objects and filtering for $\lvert b\rvert \leq 20^{\circ}$ and $\sigma_w > 9.0$ and $\sigma_p > 8.0$. The ``clean'' histogram shows all candidates after removing objects that overlap the the LMC and the Sagittarius Stream. The ``gold standard'' histogram shows all candidates that overlap with a candidate from the early EDR3 catalog. Finally, the orange histogram shows the significance of recovered known dwarfs, where we have removed known dwarfs that lie outside the parameter space probed by the mocks. The red vertical lines marks our fiducial significance threshold of $\sigma_w = 9.0$.}
\label{fig:hotspots_cut}
\end{figure}

The effect of each stage of cuts described above is shown in Figure \ref{fig:hotspots_cut}. 
The number of stars per candidate in the ``clean'' list ranges from 5 (our imposed minimum) to 462, but most candidates have around 5--10 stars (with a median of 7.0). The average magnitude of stars among the candidates is $19.4 \pm 0.5$. For candidates in the ``gold standard'' list, the median number of stars is the same (7.0), and an average magnitude of $19.4 \pm 0.4$ between candidates, consistent with the full list. The properties of candidates in our ``clean'', ``gold standard'', and ``gold standard with RR Lyrae'' lists is shown in Table \ref{tab:list_properties}.

We use the results of our search in the DES footprint to provide context for the statistics of our candidates, as we do not expect to detect any new dwarfs within that region. Within the DES footprint, there are 503 objects that pass our significance cut. We find that 83 of these candidates contain an RR Lyrae. For the $20$ known dwarf galaxies that fall within the DES footprint, six contain an associated \emph{Gaia} RR Lyrae in the field. Of these dwarfs with a \emph{Gaia} RR Lyrae, $5/6$ are recovered by the algorithm at high significance, while $1/6$ are not recovered at all. For the objects without an RR Lyrae, $3/14$ are recovered at high significance, $1/14$ is recovered at low significance, and $8/14$ are not recovered. Additionally, of the 82 candidates with an RR Lyrae, 50 correspond to known objects. This indicates that the 2 objects in the ``gold standard'' list with associated RR Lyrae have a higher likelihood of being real objects than their counterparts without \emph{Gaia}-identified RR Lyrae. However, given that $14/20$ of the DES dwarfs do not have an associated RR Lyrae star in the \emph{Gaia} catalog, we also distribute the entire ``gold standard'' candidate list for follow-up studies.

\begin{figure*}[htb!]
\centering
\includegraphics[width=\textwidth]{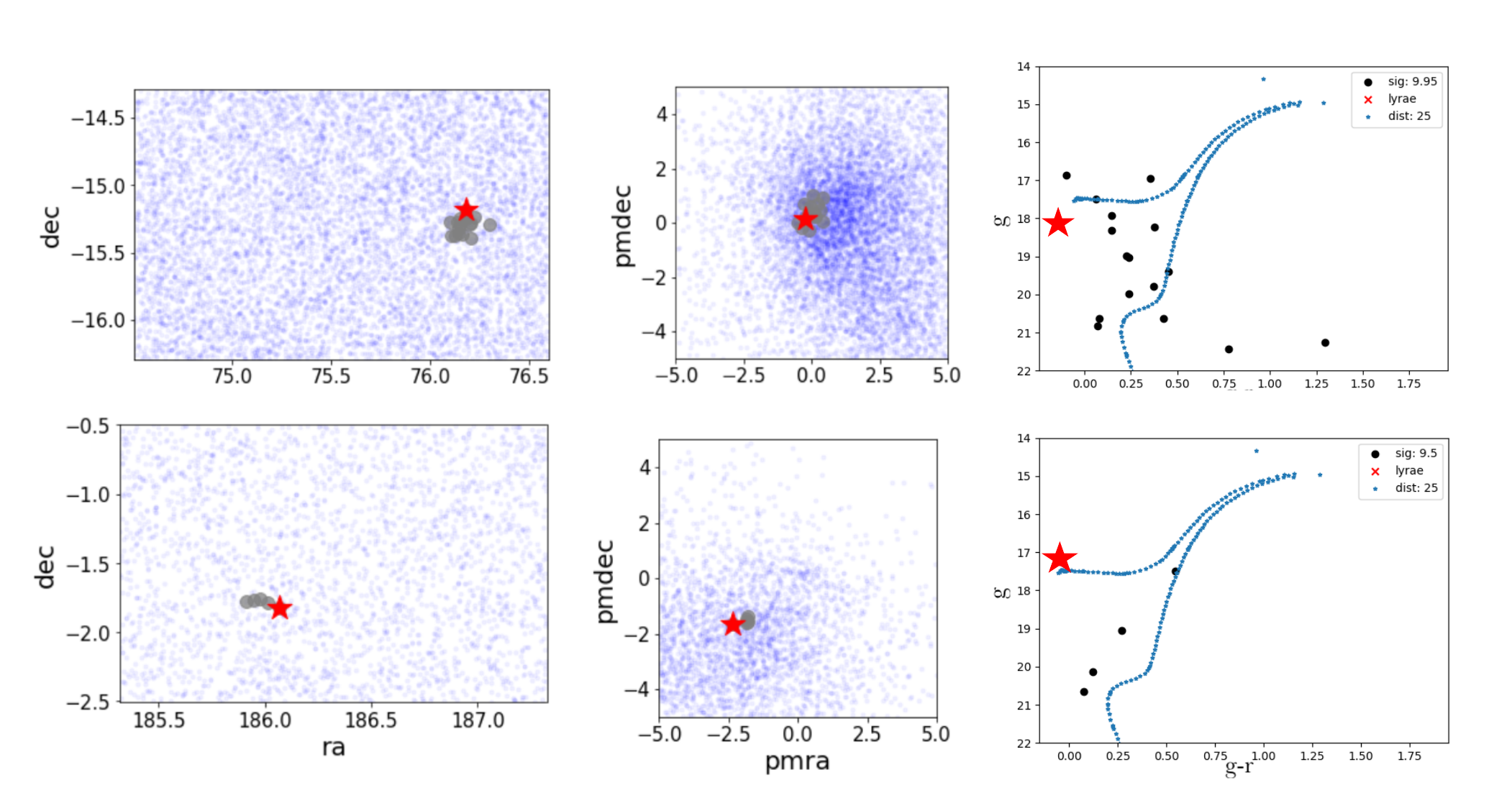}

\caption{Position (\emph{left}), proper motion (\emph{center}), and color--magnitude (\emph{right}) information for the stars in the two candidates from our ``gold standard'' list that contain associated \emph{Gaia} RR Lyrae. The stars associated with the candidate are highlighted in grey, while the associated RR Lyrae is marked as a red star in each panel. The theoretical isochrone is generated using a Padova model \citep{Bressan12084498} for a stellar population with $Z = 0.002$, an age of 12 Gyr, and at a distance of $25$ kpc, which appears to be a reasonable match to the color distribution of the candidate stars.}
\label{fig:gold_standard}
\end{figure*}

For each of the candidates in our ``gold standard with RR Lyrae'' list, we examined the candidate regions more closely. The positions of these candidates can be seen in Figure \ref{fig:dwarf_position}, while the position and proper motion of the recovered stars can be seen in Figure \ref{fig:gold_standard}, where we have chosen to highlight the associated RR Lyrae in red. Both candidates show a fairly clustered signature in both position and proper motion. We also show the color distribution of the candidate stars compared to a reference color--magnitude diagram. Both candidates are shown compared to a CMD generated using a Padova model \citep{Bressan12084498} for a stellar populations with $Z = 0.002$, an age of 12 Gyr, and at a distance of $25$ kpc, which appears to be a reasonable match to the color distribution of the candidate stars. In order to compare the \emph{Gaia} stars to a synthetic isochrone, we first convert the \emph{Gaia} filters to more standard \emph{gri} bands using a series of polynomials fit to \emph{Gaia} and DECam colors in the same field (Baumont et al.\ in prep). This is done because the \emph{Gaia} filter bands have a fairly nonstandard shape, making direct comparison to standard codes for generating isochrone models difficult. This fit is well defined between $0.3 < G_\textrm{bp} - G_\textrm{rp} < 2.0$, and nearly all of the stars we consider lie in this range. Here, we convert to DES \emph{gri} colors.

Neither of the candidates appear to be associated with a known dwarf galaxy or globular cluster. However, the first candidate is near the Orion molecular complex making dust contamination a potential issue for both \emph{Gaia} completeness in the region and the interpretation of stellar colors and magnitudes due to extinction uncertainties \citep{Rezaei2018}. The second candidate is likely associated with the Virgo Stellar Stream. The distance of the stream ($19$ kpc) broadly agrees with the color--magnitude distribution of candidate stars and the associated RR Lyrae has been previously associated with the stream \citep{Prior2009}.

The large number of ``unassociated'' candidates in the DES field also indicates that we expect many of our candidates to be spurious detections. However, the lack of precise distance information---due to very large \emph{Gaia} parallax errors at the relevant distances for the individual stars in these candidates---makes it difficult to distinguish real and spurious signals based on \emph{Gaia} data alone. One possible avenue for distinguishing real objects would be to follow up interesting candidates in deeper photometric datasets. We find that $91$ of our ``clean'' candidates fall within the DELVE survey footprint, $11$ of which are in the ``gold standard'' list. Similarly, $730$ of our ``clean'' candidates fall within the DECam legacy survey footprint, $212$ of which are in the ``gold standard'' list. We also intend to follow up convincing candidates by cross-matching with more complete RR Lyrae catalogues including the PS1 and \emph{Gaia} DR3 catalogues.

We can also distinguish real and spurious candidates by comparing the distance we estimate based on the CMD of the candidate's member stars with spectroscopic distance measurements of individual stars in the candidate. To this end, we note that the DESI Milky Way Survey \citep{Prieto_2020} is designed to obtain spectroscopic measurements of most stars between $16 < r < 19$ over 1/3 of the sky. We find of our candidates, 754/833 of the ``clean'' sample, 207/232 of the ``gold standard'' sample and 2/2 of the ``gold standard with RR Lyrae'' sample have at least one star with a \emph{Gaia} $G$-band magnitude brighter than $19$. Thus, spectroscopic measurements from the DESI survey will likely be helpful to validate the candidates of interest in future work by providing precise distance measurements. 

\begin{deluxetable*}{l c c c}
\centering
\tabletypesize{\footnotesize}
\tablecaption{\label{tab:list_properties} Summary of the properties of our three main candidate lists.}
\tablehead{ & \colhead{   Clean List   } & \colhead{   Gold Standard List (DR2xEDR3)  } &  \colhead{Gold Standard List with RR Lyrae}}
\startdata
Number of Candidates &  $833$ & $232$ & $2$ \\
Number of Candidates ($\lvert b\rvert \leq 30^{\circ}$) & $75$ & $12$ & $0$ \\
Number of Candidates (outside PS1 footprint) & $26$ & $6$ & $0$ \\
Number of Candidates (with \emph{Gaia} RR Lyrae)  & $11$ & $2$ & $2$ \\
Range in Number of Stars & $[5,462]$ & $[5,462]$ & $[5,17]$ \\
Median Number of Stars & $7$ & $7$ & -- \\
Average Stellar Magnitude (\emph{Gaia} $G$-band) & $19.4 \pm 0.5$ & $19.4 \pm 0.4$ & $18.8 \pm 0.1$ \\
\enddata
{\footnotesize}
\end{deluxetable*}

\section{Conclusions}
\label{sec:conclusion}

We have introduced a novel algorithm for detection of ultra-faint dwarf galaxies in \emph{Gaia} data. This algorithm is based on the discrete wavelet transform and has been optimized to detect dwarf galaxies as overdensities in 4D position--proper motion space. Unlike previous searches that mainly rely on spatial clustering in photometric data subject to color--magnitude filters, our algorithm leverages additional information contained in \emph{Gaia} proper motion measurements. 

We use mock dwarf galaxies injected into the \emph{Gaia} data to calibrate the algorithm's performance. Based on the mock dwarf analysis, we estimate the sensitivity of the algorithm compared to previous searches. We expect our sensitivity to be better than the PS1 search for dwarfs with $-7<M_V<-3$ at heliocentric distances of $32\ \mathrm{kpc} < D < 64\ \mathrm{kpc}$ and for dwarfs brighter than $M_V = -4$ with $64\ \mathrm{kpc} < D < 128\ \mathrm{kpc}$. Additionally, because \emph{Gaia} data covers the full sky, we are able to search over the approximately $15\%$ of the sky not previously covered by DES or PS1. We can also extend to lower latitudes than feasible in photometric searches due to the added information from the proper motion signal, which helps distinguish clusters in crowded fields. 
 
We also ran the algorithm on known Milky Way satellite galaxies, which yielded sensitivity estimates consistent with our analysis on mock dwarf galaxies. By combining these sensitivity estimates with a state-of-the-art model for the Milky Way satellite population \citep{Nadler191203303}, we estimate that our search is expected to discover $5 \pm 2$ new satellite galaxies. Four of these systems are expected to be found in the PS1 footprint and one in the region of the sky that is outside both the DES and PS1 searches.

After running the algorithm on the full \emph{Gaia} dataset and masking known objects as well as candidates likely associated with the Sagittarius Stream or Large and Small Magellenic Clouds, we recovered approximately 800 high-significance candidates located outside of the DES footprint and at $\lvert b\rvert > 20^{\circ}$. Approximately 200 of these candidates are deemed particularly promising after being cross-matched with an initial search run on the recent EDR3 data release.  Furthermore, we individually inspected the two ``gold standard'' candidates that include a \emph{Gaia}-identified RR Lyrae star.

While we see improvements in sensitivity with the \emph{Gaia} EDR3 data release, confirmation of these candidates will require follow-up in deeper photometric datasets and spectroscopic measurements if the overdensities are \emph{bona fide} dwarf galaxy candidates. We are currently working on the follow-up of some of these candidates in existing photometric datasets. We also plan to explore additional information about candidate stars provided by upcoming spectroscopic surveys, most immediately the DESI Milky Way Survey. In addition, we have made the full candidate list publicly available online for independent follow-up studies. 

In future work, we plan to leverage the flexibility of this algorithm to explore other types of phase-space structure in the Milky Way. Not only do we hope to tune the algorithm to target unknown dwarf galaxies that are difficult to access in traditional photometric searches---including low-latitude and diffuse systems---but we are also interested in extending the algorithm to search for globular clusters and stellar streams, which also present as distinct overdensities in phase space and in color--magnitude space. In addition, we plan to incorporate color--magnitude information at the search level to further enhance the purity of our candidate lists.

A particularly exciting application of our technique is the reconstruction of the fully disrupted progenitors of the Milky Way stellar halo based on the phase-space distribution of nearby stars. For example, our algorithm can potentially be optimized to detect distinct accreted stellar systems using physical priors from mock \emph{Gaia} data in hydrodynamic simulations \citep{Sanderson180610564}. This approach is complementary to analytic reconstruction techniques based on Liouville's theorem \citep{Buckley190700987} and timely given recent advances in spectroscopic measurements of the stellar halo \citep{Naidu200608625}. As demonstrated in this study, the wavelet transform's ability to distinguish structure at particular scales makes it a powerful tool in the ongoing quest to identify and understand substructure in the Milky Way.

\acknowledgments

We thank Ana Bonaca, William Cerny, and Sebastian Wagner-Carena for useful discussions, and Alex Drlica-Wagner, Marla Geha, Gregory Green, and Sidney Mau for comments on the manuscript. Additionally, we would like to thank the ApJ reviewers for helpful feedback that improved the paper. E.D-F.\ was supported by a Giddings Fellowship at Stanford. This research received support from the National Science Foundation (NSF) under grant no.\ NSF DGE-1656518 through the NSF Graduate Research Fellowship received by E.O.N., by the U.S. Department of Energy contract to SLAC no.\ DE-AC02-76SF00515, and by Stanford University.

This work has made use of data from the European Space Agency (ESA) mission
{\it Gaia} (\url{https://www.cosmos.esa.int/gaia}), processed by the {\it Gaia}
Data Processing and Analysis Consortium (DPAC,
\url{https://www.cosmos.esa.int/web/gaia/dpac/consortium}). Funding for the DPAC
has been provided by national institutions, in particular the institutions
participating in the {\it Gaia} Multilateral Agreement.

This research made use of computational resources at SLAC National Accelerator Laboratory, a U.S.\ Department of Energy Office and the Sherlock
cluster at the Stanford Research Computing Center (SRCC); the authors thank the SLAC and SRCC computing teams for their support.  This research made extensive use of \https{arXiv.org} and NASA's Astrophysics Data System for bibliographic information.

\software{
Astropy \citep{astropy},
Healpy (\http{healpy.readthedocs.io}),
Pandas \citep{pandas},
PyTorch \citep{NEURIPS2019_9015}, 
IPython \citep{ipython},
Jupyter (\http{jupyter.org}),
Matplotlib \citep{matplotlib},
NumPy \citep{numpy},
SciPy \citep{scipy},
Seaborn (\https{seaborn.pydata.org}).
}
\bibliographystyle{yahapj2}
\bibliography{references,software}

\appendix

\section{Algorithm and Hyperparameters}
\label{appendixa}

\begin{deluxetable}{l c }
\centering
\tabletypesize{\footnotesize}
\tablecaption{\label{tab:table6} Quantities used in wavelet search algorithm (Algorithms \ref{algorithm_wavelets}--\ref{algorithm_stars})}
\tablehead{\colhead{Variable} & \colhead{Value}}
\startdata
Histogram size ($H_s$) &  $96\times 96\times 96\times 96$ \\
Wavelet transform depth ($J$) & $4$ \\
Wavelets (\emph{wavelets}) &  \texttt{bior5.5} \\
Amplified scales (\emph{scales}) & $[1,2,3]$ \\
Amplification factor ($a_w$) & $100$ \\
Transform Threshold & $5\sigma$ \\
Wavelet Threshold ($\sigma_w$) & $9$ \\
Minimum number of stars ($n_{\mathrm{min}}$) & $5$ \\
\enddata
{}
\end{deluxetable}

Algorithm \ref{algorithm_wavelets} shows a schematic of the pipeline used to generate the thresholded image (e.g. third column of Figure \ref{fig:analysis_pipeline}). The description provided here is more technical than we provide in the text. We also provide a more technical description of the process of going from the thresholded image to candidate identification in Algorithm \ref{algorithm_stars}. The specific numbers chosen for this analysis are listed in Table \ref{tab:table6}. Justifications of each choice can be found in the main body of the paper. Code is publicly available at: https://github.com/edarragh/demeter

\begin{algorithm}[h!]
\SetAlgoLined
\KwData{\emph{Gaia} stars divided into equal area fields ($D$)} 
\KwResult{Transformed histogram after applying thresholding}
\DontPrintSemicolon
\;

$w_{\mathrm{field}} =  0.0175$\;
$h_{\mathrm{field}}  =  0.0175$\; 
$D = \texttt{to\_gnomonic}(D, w_{\mathrm{field}}, h_{\mathrm{field}})$\footnote{Equation \ref{eq:gnomx}, \ref{eq:gnomy}}\;

\;
\SetKwFunction{FMain}{generate\_random}
\SetKwProg{Fn}{Function}{:}{\KwRet [$\alpha_{2000_{\mathrm{rand}}}$,~$\delta_{2000_{\mathrm{rand}}}$,~$\mu_{\alpha_{\mathrm{rand}}}$,~$\mu_{\delta_{\mathrm{rand}}}$]\;}
\Fn{\FMain{D, $w_{\mathrm{field}}$, $h_{\mathrm{field}}$}}{
        $n = \texttt{poisson}(n_{stars}(D))$ \;
        $\alpha_{2000_{\mathrm{rand}}}$, $\delta_{2000_{\mathrm{rand}}}$ = \texttt{uniform}($n$, $w_{\mathrm{field}}$), \texttt{uniform}($n$, $h_{\mathrm{field}}$)\;
        \For{$i$ in \texttt{range}(4)\;}{$GMM_i$ = \texttt{gmm}(i, \texttt{zip}($D$$[\mu_{\alpha}]$, $D$$[\mu_{\delta}]))$\; bic$_i$ = bic($GMM_i$) \;} 
        N = \texttt{min}(bic$_i$) \;
        $\mu_{\alpha_{\mathrm{rand}}}$, $\mu_{\delta_{\mathrm{rand}}}$ = $GMM_N$.\texttt{sample}($n$) \; }
$R$ = \texttt{generate\_random}($D$, $w_{\mathrm{field}}$, $h_{\mathrm{field}}$)\;   
$H$ = \texttt{histogram(}$D$, $H$$_s$)\;
$H$$_{\mathrm{rand}}$ = \texttt{histogram}($R$, $H$$_s$)\;
\;

\DontPrintSemicolon
\SetKwFunction{Wave}{wavelet\_transform}
\SetKwProg{Pn}{Function}{:}{}
\Pn{\Wave{H}}{
        maps = []\;
        \For{$w$ \textrm{ \upshape in} wavelets\;}{$c_H$, $d_H$ = $w$($H$, $J$)\; \For{j \textrm{\upshape in} scales \;}{$d_j = a_w \times d_j$ \;} $m$ = $w^{-1}([c_H$, $d_H$], $J$) \; maps.\texttt{add}($m$)}\;
        \KwRet \texttt{sum}(maps)\;
  }
  \; 
 
 $HT$ = \texttt{smooth(wavelet\_transform}($H$), 3) \;
 $HT_{\mathrm{rand}}$ = \texttt{smooth(wavelet\_transform}($H_{\mathrm{rand}}), 3$) \;

$HT$' = \texttt{standardize}($HT$, $HT_{\mathrm{rand}}$)\footnote{Equation \ref{eq:sigmaw}}\; 

$HT$' = \texttt{threshold}($HT$')\footnote{set $p$' $< 5$ to zero} \; 
\KwRet{\textrm{ \upshape $HT$'}}\; 
 \caption{Wavelet Algorithm Pipeline}
  \label{algorithm_wavelets}

\end{algorithm}

\begin{algorithm}[h!]
\SetAlgoLined
\KwData{Transformed 4D histogram ($HT$') after thresholding has been applied} 
\KwResult{Select candidate stars from \emph{Gaia} field ($D$)}
\DontPrintSemicolon
\;

neighbors = \texttt{ones}((3,3,3,3)) \;

$HT$'$_{\mathrm{labeled}}$, $n$ = \texttt{label}(\textrm{ \upshape $HT$', structure = neighbors}) \; 
blobs = [] 

\For{k \textrm{ \upshape in range (}n\textrm{\upshape)} \;}{mask = $HT$'$_{\mathrm{labeled}}$ == $k$ \; \If{\textrm{ \upshape\texttt{sig}($HT$'}$_{\mathrm{labeled}}$\textrm{ \upshape[mask]) > 9.0} \;}{$\alpha_{2000}$, $\delta_{2000}$, $\mu_{\alpha}$, $\mu_{\delta}$, \textit{r}$_{\alpha_{2000}}$, \textit{r}$_{\delta_{2000}}$, \textit{r}$_{pm}$ = \texttt{gmm}(1, \textrm{ \upshape $HT$'}$_{\mathrm{labeled}}$)\footnote{Equation \ref{eq: radii}}\; blobs.\texttt{append}[$\alpha_{2000}$, $\delta_{2000}$, $\mu_{\alpha}$, $\mu_{\delta}$, \textit{r}$_{\alpha_{2000}}$, \textit{r}$_{\delta_{2000}}$, \textit{r}$_{pm}$] \;}}
candidates = [] \;
\For{\textrm{\upshape blob in blobs}}{mask = $D$.\texttt{isin}(blob) \; stars = $D$[mask]\; \If{\textrm{ \upshape \texttt{len}(stars) > $n_{\mathrm{min}}$}\;}{candidates.\texttt{append}(stars)\;}}
\KwRet{\textrm{ \upshape candidates}}\; 
 \caption{Star Selection Pipeline}
 \label{algorithm_stars}
\end{algorithm}

\end{document}